\documentclass[12pt]{article}
\usepackage[utf8]{inputenc}
\usepackage{geometry}
\usepackage{fullpage}

\usepackage{float}
\usepackage{amsmath}
\usepackage{amssymb}
\usepackage{esint}
\usepackage{hyperref}

\makeatletter
\@ifundefined{date}{}{\date{}}
\pdfoutput=1
\usepackage{graphicx}
\usepackage{color}

\newcommand{\f}[2]{\frac{#1}{#2}}
\newcommand{\cor}[1]{\left\langle{#1}\right\rangle}
\newcommand{\corr}[3]{\bra{#1} {#2} \ket{#3}}
\newcommand{\ket}[1]{\left|{#1}\right\rangle}
\newcommand{\bra}[1]{\left\langle{#1}\right|}

\newcommand{\eq}{\begin{equation}}
\newcommand{\eqx}{\end{equation}}
\newcommand{\eqn}{\begin{eqnarray}}
\newcommand{\eqnx}{\end{eqnarray}}

\renewcommand{\th}{\theta}

\newcommand{\al}{\alpha}
\newcommand{\alb}{\bar{\alpha}}

\newcommand{\eps}{\varepsilon}

\newcommand{\vac}{\emptyset}

\newcommand{\XX}{{\mathcal X}}
\newcommand{\FF}{{\mathcal F}}
\newcommand{\OO}{{\mathcal O}}

\newcommand{\oct}{{\mathcal O}^{\th_1,\th_2}}
\newcommand{\neum}{{\mathcal N}^{\th_1,\th_2}_{L_1}}

\title{From the octagon to the SFT vertex - gluing and multiple wrapping}

\author{Zoltan Bajnok$^{a}$\thanks{e-mail: {\tt bajnok.zoltan@wigner.mta.hu}},\ \  
Romuald A. Janik$^{b}$\thanks{e-mail: {\tt romuald@th.if.uj.edu.pl}} \\ \\ 
\small 
${}^a$ MTA Lend\"ulet Holographic QFT Group\\\small
Wigner Research Centre for Physics\\\small
H-1525 Budapest 114, P.O.B. 49, Hungary\\\small
${}^b$ Institute of Physics\\\small
Jagiellonian University\\\small
ul. {\L}ojasiewicza 11, 30-348 Krak{\'o}w, Poland}

\makeatother

\begin{document}
\maketitle
\begin{abstract}
We compare various ways of decomposing and decompactifying the string field theory vertex
and analyze the relations between them. We formulate axioms for the octagon and show
how it can be glued to reproduce the decompactified pp-wave SFT vertex
which in turn can be glued to recover the exact finite volume pp-wave Neumann
coefficients.
The gluing is performed by resumming multiple wrapping corrections.
We observe important nontrivial contributions at the multiple
wrapping level which are crucial for obtaining the exact results.
\end{abstract}
\vfill{}

\pagebreak{}

\section{Introduction}

The AdS/CFT correspondence relates string theories on anti de Sitter backgrounds to 
conformal gauge theories on the boundary of these spaces \cite{adscft1}. The energies of  string states correspond 
to the scaling dimensions of local gauge invariant operators which determine  the space time dependence of the conformal 
2- and 3-point functions  completely. In order to
build all higher point correlation functions of the CFT one needs to determine the  3-point 
couplings, which is in the focus of recent research. 

String theories on many AdS backgrounds are integrable \cite{Bena:2003wd,Arutyunov:2008if,Stefanski:2008ik,Babichenko:2009dk}  and this 
 miraculous infinite symmetry is the one which enables us to solve the quantum string theory dual to the  strongly 
coupled gauge theory \cite{intreview}.  In the prototypical example the type I\!\,IB superstring theory on the 
$AdS_5\times S^5$ background  is dual to the maximally supersymmetric 4D gauge theory.
 Integrability shows up in the planar limit and interpolates between the weak and strong coupling sides. 
The spectrum of string theory, i.e. the scaling dimensions of  local gauge-invariant operators are 
mapped to the finite volume spectrum of the integrable theory, which has been determined by adapting 
finite size techniques such as Thermodynamic Bethe Ansatz \cite{TBA1,TBA2,TBA3}  
(consequently developed into a NLIE \cite{BH} and the quantum spectral curve \cite{QSC1,QSC2}). 
Further important observables such as 3-point correlation functions or nonplanar corrections 
to the dilatation operator are related to string interactions. A generic approach to the string 
field theory vertex was introduced in \cite{SFT} 
which can be understood as a sort of finite volume form factor of non-local operator 
insertions in the integrable worldsheet theory. There is actually one case when the 3-point function corresponds to a 
form factor of a \emph{local} operator insertion. In the case of heavy-heavy-light
operators the string worldsheet degenerates into a cylinder and the SFT vertex is nothing but a diagonal finite volume form factor, see 
\cite{Bajnok:2014sza,Hollo:2015cda,Bajnok:2016xxu}.
Another approach through cutting the string worldsheet corresponding to a 3-point 
correlation function into two hexagons was introduced in \cite{HEXAGON},  see also \cite{Eden:2015ija,Basso:2015eqa,Jiang:2016ulr,Jiang:2015bvm,Jiang:2016dsr,Basso:2017muf} for further developments.

The string field theory vertex describes a process in which a big
string splits into two smaller ones. In light-cone gauge fixed string
sigma models on $AdS_5\times S^5$ and some similar backgrounds, the string worldsheet
theory is integrable and the conserved J-charge serves as the volume,
so that the size\footnote{In this paper we will use terms size and volume
interchangeably to mean the circumference of the cylinder on which the worldsheet QFT
of the string is defined.} of the incoming string exactly equals the sum of the sizes
of the two outgoing strings.
Initial and final states are characterized
as multiparticle states of the worldsheet theory on the respective cylinders
and we are interested in the asymptotic time
evolution amplitudes, which can be essentially described as finite volume form
factors of a non-local operator insertion representing the emission of the third string. 
In order to be able to obtain functional equations for these
quantities we suggested in \cite{SFT} to analyze the decompactification
limit, in which the incoming and one outgoing volume are sent to infinity,
such that their difference is kept fixed. 
We called this quantity the decompactified string field
theory (DSFT) vertex or decompactified Neumann coefficient. We formulated
axioms for such form factors, which depend explicitly on the size
of the small string, and determined the relevant solutions in the
free boson (plane-wave limit) theory. Taking a natural Ansatz for
the two particle form factors we separated the kinematical and the
dynamical part of the amplitude and determined the kinematical Neumann
coefficient in the AdS/CFT case \cite{KINEMATICAL}, too. These solutions automatically
contain all wrapping corrections in the remaining finite size string,
which makes it very difficult to calculate them explicitly in the interacting case, especially
for more than two particles. It is then
natural to send the remaining volume to infinity and calculate the
so obtained \emph{octagon} amplitudes. One can go even further and
introduce another cut between the front and back sheets leading to
two hexagons, which were introduced and explicitly calculated in \cite{HEXAGON}. Since we
are eventually interested in the string field theory vertex, we have to
understand how to glue back the cut pieces. This paper is an attempt
going into this direction. Clearly, gluing two hexagons together we should
recover the octagon amplitude\footnote{We will not consider this case, however, in the current paper.}. 
Gluing two edges of the octagon we
get the DSFT vertex, while gluing the remaining two edges we would obtain
the \emph{finite} volume SFT vertex, which would be the ultimate goal
for the interacting theory. 

The study of various observables in integrable quantum field theories
in\emph{finite} volume in a natural way can be decomposed into a
number of stages. Firstly, the problem posed in infinite volume typically
yields a set of axioms or functional equations for the observable
in question which often can be solved explicitly. The key property
of the infinite volume formulation is the existence of analyticity
and crossing relations which allow typically for formulating functional
equations \cite{Zamolodchikov:1978xm,Mussardo:1992uc,KW,Smirnov}. 
Secondly one considers the same problem in a large finite
volume neglecting exponential corrections of order $e^{-mL}$. In
this case the answers are mostly known like for the energy levels,
generic form factors\footnote{By this we mean form factors with no coinciding rapidities in any
channel.} and diagonal form factors \cite{Pozsgay:2007kn,Pozsgay:2007gx}. 
However, some of these answers are still
conjectural and are not known in various interesting cases. Thirdly,
one should incorporate the exponential corrections of order $e^{-mL}$,
which are often termed as wrapping corrections as they have the physical
interpretation of a virtual particle wrapping around a noncontractible
cycle. The key example here are the L\"uscher corrections for the mass
of a single particle \cite{LUSCHER} and their multiparticle generalization \cite{KONISHI}. Once
one wants to incorporate multiple wrapping corrections, the situation
becomes much more complicated however in some cases this can be done
\cite{BOMBARDELLI}.

In the case of the spectrum of the theory on a cylinder, fortunately
one does not need to go through the latter computations as there exists
a Thermodynamic Bethe Ansatz formulation which at once resums automatically
all multiple wrapping corrections and provides an exact finite volume answer \cite{Zamolodchikov}. Unfortunately
for other observables like the string interaction vertex we do not
have this technique at our disposal and we may hope that understanding
the structure of multiple wrapping corrections will shed some light
on an ultimate TBA like formulation. This is another motivation for the present work.
In fact one of the new results of the present paper is an integral representation
for the exact pp-wave Neumann coefficient which involves a measure factor 
reminiscent of various TBA formulas.

In \cite{HEXAGON}, a formula for gluing two hexagons was proposed: insert a complete basis of particles
on the mirror edge\footnote{In the relativistic case this would correspond to inserting particles
in the channel with space and time interchanged i.e. with rapidities
$\th+i\pi/2$.} and sum over them 
\begin{equation}
\label{e.glue}
\sum_{n=0}^{\infty}\f{1}{n!}\int\mu_{n}e^{-\sum_i E_{i}L}
\end{equation}
This is in fact a rather formal expression as the observable in question
is divergent. Also we allowed for a generic measure factor. It will indeed turn out
that the measure factor is nontrivial for multiple wrapping.
In this paper we analyze the multiple wrapping terms for a massive free boson theory
which corresponds to the relevant quantities being evaluated for the pp-wave
geometry. 

The outline of this paper is as follows. In section 2 we will review the
decompactified SFT vertex axioms as well as introduce the axioms for the
octagon. We will also deduce the measure factor by requiring that
gluing the octagon through (\ref{e.glue}) reproduces the exact pp-wave
decompactified SFT vertex.
Then in section 3, we will revisit the gluing procedure (\ref{e.glue})
from the point of view of cluster expansion (or equivalently compactification
in the mirror channel) and isolate the key ingredients which are necessary
for obtaining the finite volume answer for a generic observable. 
We will also illustrate this structure with the well known relativistic
examples of ground state energy and LeClair-Mussardo formula for the 
finite volume 1-point expectation value \cite{Leclair:1999ys}.
In the following two sections we will show that one can provide natural choices
for these ingredients which enable us to glue the octagon into the
decompactified SFT vertex and then glue the decompactified SFT vertex
into the exact finite volume pp-wave vertex\footnote{Here we are concerned with
just the bosonic case so we do not consider issues related with the prefactor.}. 
We close the paper with a discussion and two appendices, one of which contains
the derivation of the integral representation of the pp-wave Neumann coefficient,
and the other a discussion of the relation between octagon axioms and decompactified
SFT vertex axioms in the context of the gluing formula.

\section{Cutting pants into DSFT vertex, octagon}

\begin{figure}
\begin{centering}
\includegraphics[height=3cm]{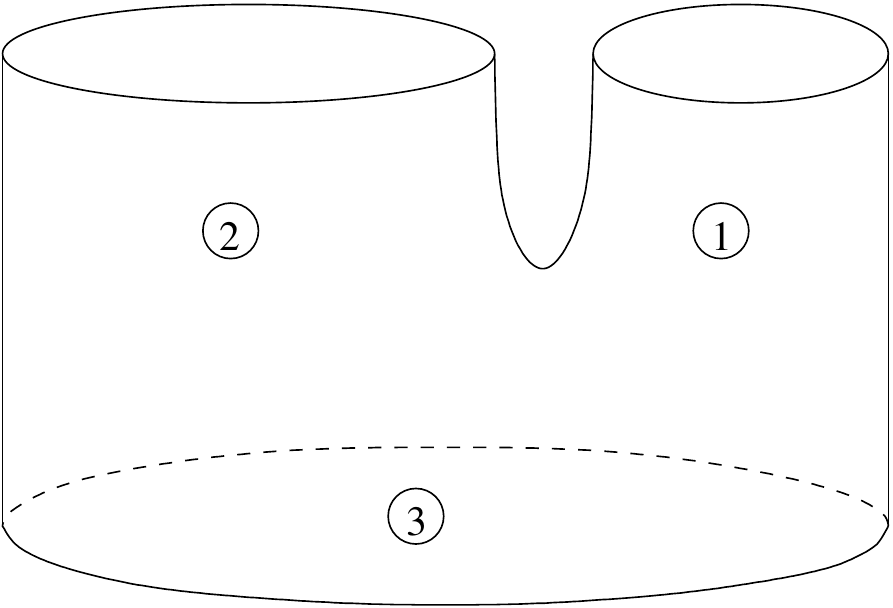}~~~~~\includegraphics[height=3cm]{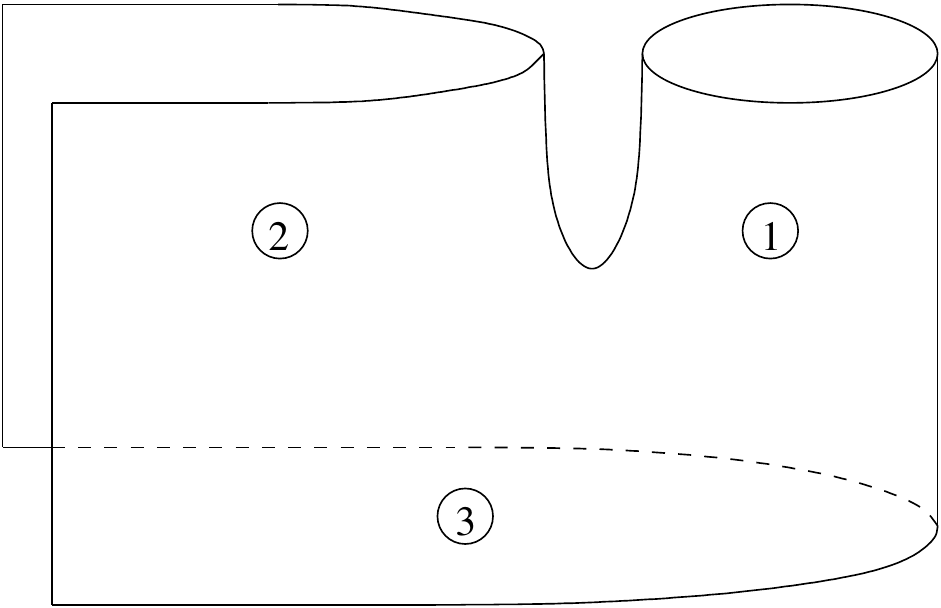}~~~~~\includegraphics[height=3cm]{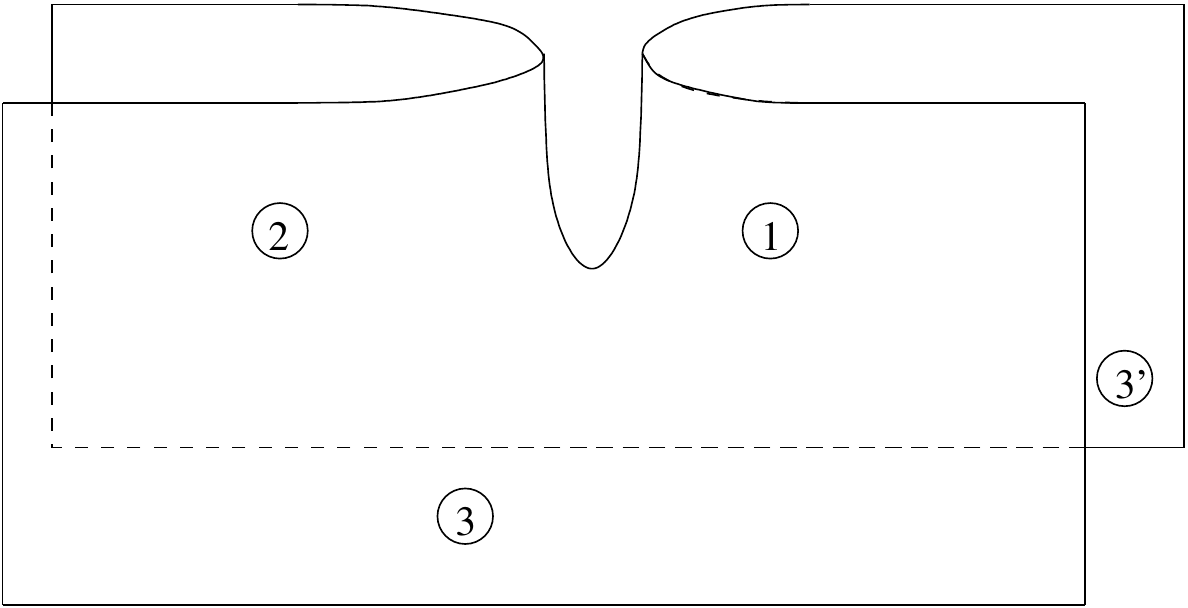}
\par\end{centering}

\caption{Splitting of a big string into two smaller ones and its decompactified
versions.}

\label{SFTV}
\end{figure}

The string field theory vertex describes the amplitude of the process
in which a big string (\#3) splits into two smaller ones (\#1 and
\#2), see left of Figure \ref{SFTV}. In light-cone gauge fixed string
sigma models the conserved J-charge serves as the volume, which adds
up in the process $J_{3}=J_{2}+J_{1}$. Initial and final states are
characterized as finite volume multiparticle states and the asymptotic
time evolution amplitudes can be understood as finite volume form factors of a non-local
operator insertion. In calculating these quantities we go to the decompactification
limit, in which the volumes $J_{3}$ and $J_{2}$ are sent to infinity,
such that their difference $J_{3}-J_{2}=L$, which is the size of the remaining
closed string, is kept fixed leading
to infinite volume form factors, see the middle of Figure \ref{SFTV}.
These form factors automatically contain all wrapping corrections
in the remaining finite size string, which makes very difficult to
calculate them explicitly. It is then natural to send the remaining
volume to infinity and calculate the so obtained \emph{octagon} amplitudes.
See the right of Figure \ref{SFTV} for the geometry. Since eventually
we are interested in the string field theory vertex we have to glue
back the cut edges. Gluing two edges of the octagon we get the decompactified
SFT vertex, while gluing the remaining two edges we obtain the seeked
for SFT vertex. 

\begin{figure}
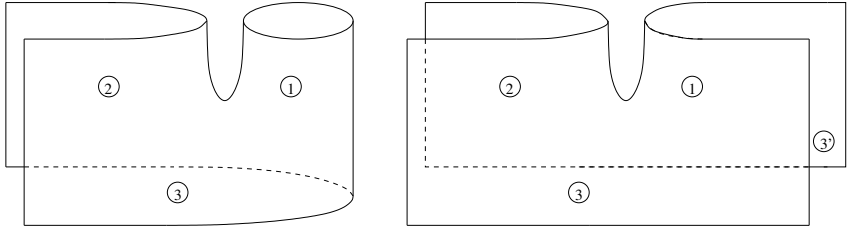

\begin{centering}
\includegraphics[height=3cm]{sft2.pdf}~~~~~\includegraphics[height=3cm]{sft4.pdf}
\par\end{centering}

\caption{The decompactified SFT vertex on the left and the octagon, on the
right. }

\label{OCT}
\end{figure}

\subsection{The decompactified SFT vertex}

In our previous paper \cite{SFT} we formulated the axioms of the DSFT vertex,
which we also called the generalized Neumann coefficients. 
Here for simplicity we quote the axioms for a relativistic theory
with a single type of particle.
For initial
particles living on string \#2 with rapidities $\theta_{i}$ they
read as follows: 
\begin{itemize}
\item The exchange axiom is 
\begin{equation}
N_{L}(\theta_{1},\dots,\theta_{i}, \theta_{i+1},\dots,\theta_{n})=S(\theta_{i}-\theta_{i+1})
N_{L}(\theta_{1},\dots,\theta_{i+1}, \theta_{i} \dots,\theta_{n})
\end{equation}
\item The periodicity axiom explicitly includes the volume of the small emitted
string: 
\begin{equation}
N_{L}(\theta_{1}+2i\pi,\theta_{2},\dots,\theta_{n})=
e^{ip_{1}L}N_{L}(\theta_{2},\dots,\theta_{n},\theta_{1})
\end{equation}

\item The kinematical singularity axiom, which relates form
factors with different particle numbers, takes the form:
\begin{equation}
-i\mbox{Res}_{\theta'=\theta}N_{L}(\theta' +i\pi,\theta,\theta_{1},\dots,\theta_{n})=(1-e^{-ipL}\prod_{j}S(\theta-\theta_{j}))N_{L}(\theta_{1},\dots,\theta_{n})
\end{equation}

\end{itemize}
We have determined in \cite{SFT} the 2-particle solution for the free
boson theory $S(\theta)=1$, which reads as\footnote{Here we choose the $ N_L(\emptyset)=1$ 
normalization.} 
\begin{equation}
N_{L}(\theta_{1},\theta_{2})=-\frac{1}{2\cosh\frac{\theta_{1}-\theta_{2}}{2}}d_{L}(\theta_{1})d_{L}(\theta_{2})
\end{equation}
where the functions $d_{L}(\th)$ involve all order wrapping terms.
They are given explicitly in terms of deformed Gamma functions which
have a rather nontransparent definition. 
The above formula exactly coincides with the decompactification limit
of the pp-wave Neumann coefficient \cite{LSNS}.
Remarkably enough, there
exists a very compact and transparent integral formula for $d_{L}(\th)$
which we derive in Appendix~A. It takes the form 
\begin{equation}
d_{L}(\th)=e^{-\int_{-\infty}^{\infty}\f{du}{2\pi}k(u-\theta)\log(1-e^{-mL\cosh u})}\quad;\qquad k(\theta)=-\frac{1}{\cosh(\th)}
\end{equation}
The multiparticle solutions can be fixed from the kinematical residue
equation and have the form 
\begin{equation}
N_{L}(\theta_{1},\dots,\theta_{n})=\sum_{\mathrm{pairings}}\prod_{(i,j)\,\mathrm{pairs}}N_{L}(\theta_{i},\theta_{j})
\end{equation}
Thus we sum for all possible pairings of the rapidities $\{\theta_{i}\}$
and take the product for the pairs of the 2-particle expressions.
Clearly this form is compatible with Wick theorem in the free boson
theory. 

\begin{figure}
\begin{centering}
\includegraphics[height=3cm]{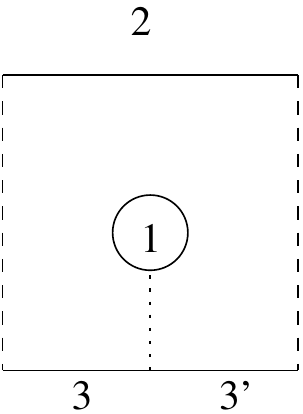} ~~~~~~~\includegraphics[height=4cm]{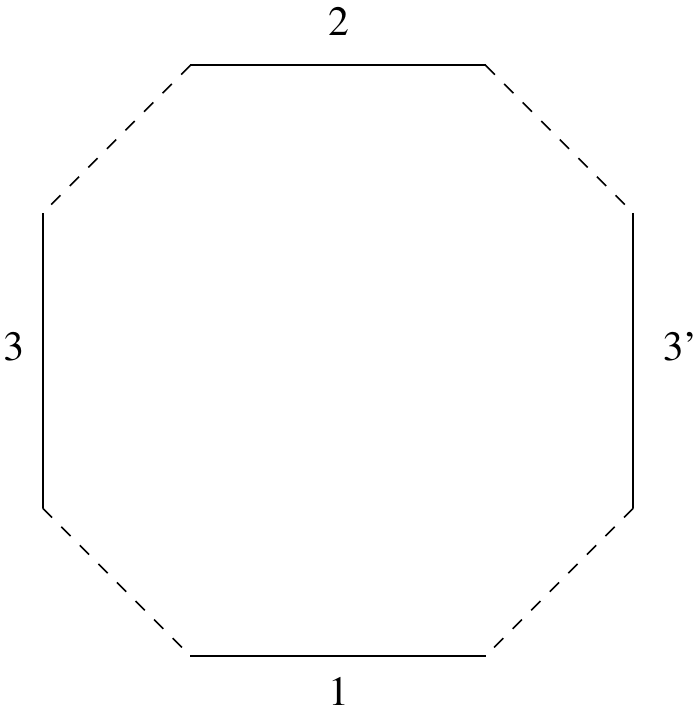}
\par\end{centering}
\caption{The kinematical domains of the DSFT vertex is on the
left, while that of the octagon amplitude is on the right. The glued mirror edges between \#31 and \#3'1 are 
indicated by a dotted line. The finite size string represented by the circle serves as a non-local operator 
insertion in the square topology.}
\label{glue_octagon}
\end{figure}

From the decompactified string vertex one can go in two opposite directions.
Either one can glue together the remaining two mirror edges (the dashed lines between \#2 and \#3 and
between \#2 and \#3' in Figure~\ref{glue_octagon}) thus obtaining the finite size
SFT vertex, which is really the ultimate goal of this program, or one can go in
the opposite direction and send the remaining volume $L$ to infinity thus
obtaining the octagon.

In the case of the free massive boson (the pp-wave) the exact finite size
SFT vertex Neumann coefficient (up to an overall normalization) can be expressed very compactly as
\begin{equation}
N_{L_{1}}^{L_{2}}(\th_{1},\th_{2})=N_{L_{1}}(\th_{1},\th_{2})
\cdot\f{d_{L_{2}}(\th_{1})}{d_{L_{3}}(\th_{1})}\cdot\f{d_{L_{2}}(\th_{2})}{d_{L_{3}}(\th_{2})}
\end{equation}
In section~5 we will describe how this form can be obtained by gluing together the decompactified
SFT vertex.

Now, however, we will concentrate on the octagon which appears when we send the remaining volume,
$L$, to infinity. Effectively, this limit not only sends the volume
of string \#1 to infinity but also cuts the space of string \#3 into
two disconnected pieces, which we denote by \#3 and \#3'. They are
connected by crossing through string \#1 on one side and through string
\#2 on the other. This suggests the octagon description as shown of
Figure \ref{glue_octagon}. Let us formulate the functional relations
for this quantity. 

\subsection{The octagon}

The octagon amplitude, when particles with rapidities $\theta_{i}$
are in string \#2, satisfy the following axioms: 
\begin{itemize}
\item The exchange axiom relates particles on the same kinematical edge
to each other thus is not changed compared to the DSFT vertex axioms:
\begin{equation}
O(\theta_{1},\dots,\theta_{i},\theta_{i+1},\dots,\theta_{n})=S(\theta_{i}-\theta_{i+1})O(\theta_{1},\dots,\theta_{i+1},\theta_{i},\dots,\theta_{n})
\end{equation}

\item In the periodicity properties we have to cross a particle from domain \#2 to
\#3 first, then from \#3 to \#1, then from \#1 back to
\#3' and finally to \#2 leading to a $4i\pi$ periodicity:
\begin{equation}
O(\theta_{1}+4i\pi,\theta_{2},\dots,\theta_{n})=O(\theta_{2},\dots,\theta_{n},\theta_{1})
\end{equation}

\item In the kinematical singularity axiom particles in domain \#2 can
feel particles in domain \#3 only by crossing with $i\pi$, (and
not by crossing with $-i\pi$ ), thus we have
\begin{equation}
-i\mbox{Res}_{\theta'=\theta}O(\theta'+i\pi,\theta,\theta_{1},\dots,\theta_{n})=O(\theta_{1},\dots,\theta_{n})
\end{equation}
i.e no S-matrix factors appear, which make their determination easier. 
\end{itemize}
Particles on different edges of the octagon can be obtained by analytical
continuation, what we describe in detail in the Appendix B.

The two particle octagon solution for the free boson theory is 
\begin{equation}
O(\theta_{1},\theta_{2})=-\frac{1}{2\cosh\frac{\theta_{1}-\theta_{2}}{2}}
\end{equation}
The multiparticle solutions can be fixed from the kinematical singularity
axiom and take the form 
\begin{equation}
O(\theta_{1},\dots,\theta_{n})=\sum_{\mathrm{pairings}}\prod_{(i,j)\,\mathrm{pairs}}O(\theta_{i},\theta_{j})
\end{equation}
Our main problem now is to understand how to obtain the DSFT vertex with
string \#1 having a finite size $L$, by gluing together the two mirror
edges between \#1 and \#3 and between \#1 and \#3' (see figure \ref{glue_octagon}).

\subsection{Naive resummation of the octagon}

\label{s.naiveoctagon}

\begin{figure}[h]
\begin{centering}
\includegraphics[height=3.5cm]{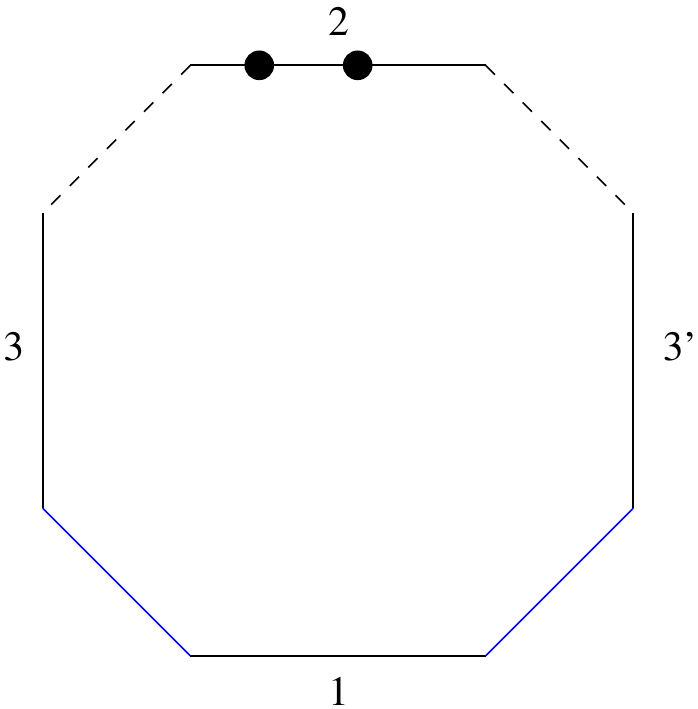}$=$\includegraphics[height=3.5cm]{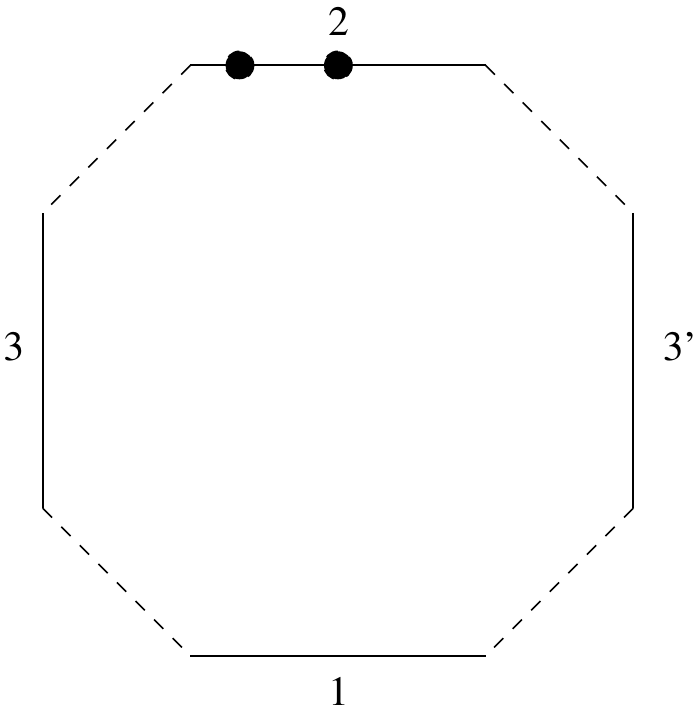}$+$\includegraphics[height=3.5cm]{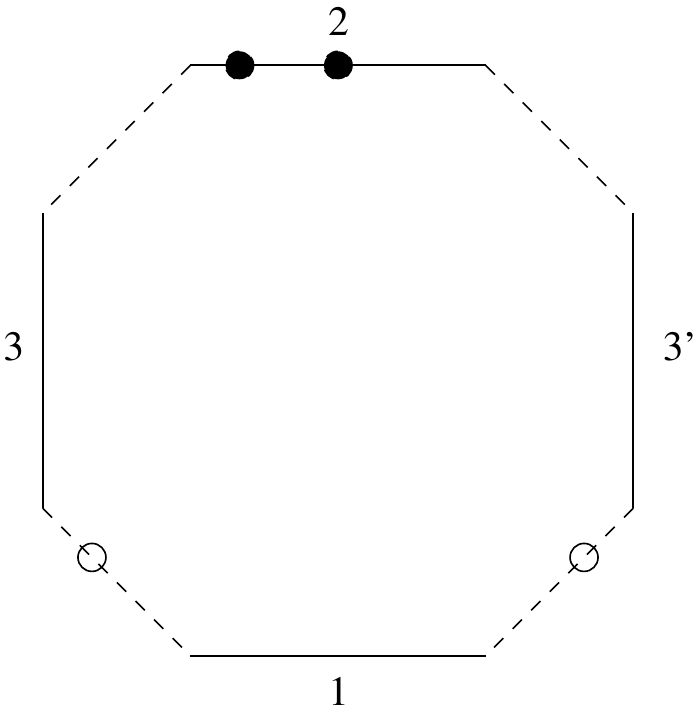}$+$\includegraphics[height=3.5cm]{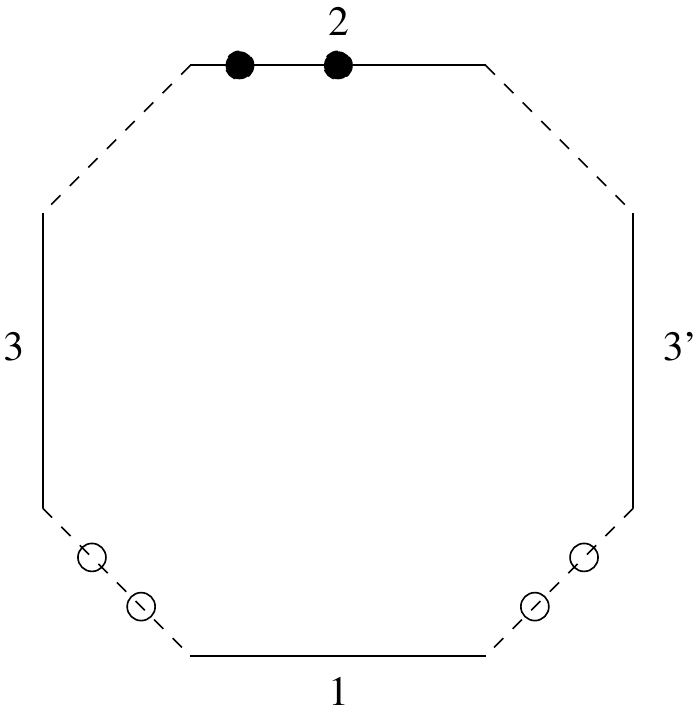}$+\dots$
\par\end{centering}
\caption{Summing up octagons to get the DSFT vertex. The blue edges are glued
together by summing up for all multi-particle mirror states, represented by empty circles. 
Physical particles are represented by solid circles. }
\label{glu_2pt_oct}
\end{figure}

\begin{figure}[h]
\begin{centering}
\includegraphics[width=3cm]{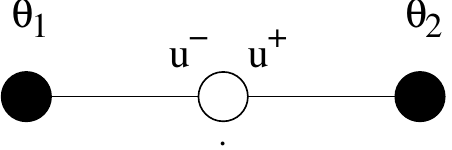} \hspace{1cm} \includegraphics[width=3cm]{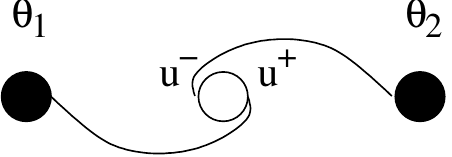}  
\hspace{1cm} \includegraphics[width=3cm]{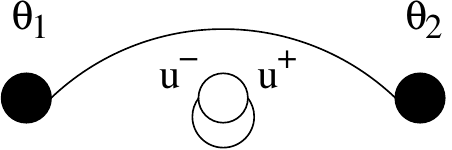}
\par\end{centering}
\caption{$O(\theta_1,\theta_2,u^+,u^-)$ in graphical notation. The first two diagrams are connected, while the last is disconnected. }
\label{rules}
\end{figure}

A very formal definition of gluing two mirror edges was proposed in
\cite{HEXAGON}. We demonstrate this idea on the example how the DSFT could
be obtained from the resummation of octagons. The idea of the gluing
is to interpret the cutting as a resolution of the identity 
\begin{equation}
\sum_{n=0}^{\infty}\frac{1}{n!}\prod_{i=1}^{n}\int_{-\infty}^{\infty}\frac{du_{i}}{2\pi}\mu(\{u_{i}\})e^{-\sum_{i}E(u_{i})L}\vert u_{1},\dots,u_{n}\rangle\langle u_{n},\dots,u_{1}\vert
\end{equation}
where $\vert u_{1},\dots,u_{n}\rangle$ denotes an infinite volume
mirror state living between the spaces \#3 and \#1, while $\langle u_{n},\dots,u_{1}\vert$
is its dual mirror space living between the space \#1 and \#3'.
In formulas it means for a two particle DSFT vertex that
\begin{eqnarray}
N_{L}(\theta_{1},\theta_{2}) & =\frac{1}{\mbox{norm}}\biggl\{ & O(\theta_{1},\theta_{2})+\int_{-\infty}^{\infty}\frac{du}{2\pi}\mu_{1}(u)O(\theta_{1},\theta_{2},u^{+},u^{-})e^{-LE(u)}+\\
 &  & \hspace{-1.5cm}\frac{1}{2}\int_{-\infty}^{\infty}\frac{du_{1}}{2\pi}\int_{-\infty}^{\infty}\frac{du_{2}}{2\pi}\,\mu_{2}(u_{1},u_{2})O(\theta_{1},\theta_{2},u_{1}^{+},u_{2}^{+},u_{2}^{-},u_{1}^{-})e^{-L\left(E(u_{1})+E(u_{2})\right)}+\dots\biggr\}\nonumber 
\label{formal}
\end{eqnarray}
where $u^{\pm}=u\pm\frac{i3\pi}{2}$. Graphically it can be represented
as on Figure \ref{glu_2pt_oct}. 
Since the mirror particle-anti-particle pairs come on the opposite
edges of the octagon the amplitude is singular due to the kinematical
singularity axioms.However, it is very natural to normalize the amplitude by the
``infinite'' empty glued octagon: 
\begin{eqnarray}
\textrm{norm} & = & 1+\int_{-\infty}^{\infty}\frac{du}{2\pi}\mu_{1}(u)O(u^{+},u^{-})e^{-LE(u)}+\\
 &  & \frac{1}{2}\int_{-\infty}^{\infty}\frac{du_{1}}{2\pi}\int_{-\infty}^{\infty}\frac{du_{2}}{2\pi}\,\mu_{2}(u_{1},u_{2})O(u_{1}^{+},u_{2}^{+},u_{2}^{-},u_{1}^{-})e^{-L\left(E(u_{1})+E(u_{2})\right)}+\dots\nonumber 
\end{eqnarray}
which exactly suffers from the same divergences. Indeed, 
the special "free" form of the octagon amplitudes guarantees that the normalization
in the denominator removes all the disconnected singular terms and
only finite regular expressions remain: 
\begin{eqnarray}
N_{L}(\theta_{1},\theta_{2}) & = & O(\theta_{1},\theta_{2})+\int_{-\infty}^{\infty}\frac{du}{2\pi}\mu_{1}(u)O^c(\theta_{1},\theta_{2},u^{+},u^{-})e^{-LE(u)}+\\
 &  & \frac{1}{2}\int_{-\infty}^{\infty}\frac{du_{1}}{2\pi}\int_{-\infty}^{\infty}\frac{du_{2}}{2\pi}\,\mu_{2}(u_{1},u_{2})O^c(\theta_{1},\theta_{2},u_{1}^{+},u_{2}^{+},u_{2}^{-},u_{1}^{-})e^{-L\left(E(u_{1})+E(u_{2})\right)}+\dots\biggr\}\nonumber 
\end{eqnarray}
where $O(\theta_{1},\theta_{2},u_{1}^{+},u_{1}^{-},\dots,u_{n}^{+},u_{n}^{-})_{c}$
denotes the connected part, i.e. the one which is connected with the
following graphical rules: Put the first vertex for $\theta_{1}$
and the last for $\theta_{2}$, while in between $n$ vertices for
each $u_{i}$. Left side of the $u_{i}$ represents $u_{i}^{-}$,
while the right $u_{i}^{+}.$ For each propagator $O(\theta_{1},u_{j}^{\pm})$
draw an edge from $\theta_{1}$ to the right/left of $u_{j}$. For
a propagator $O(u_{j}^{\epsilon_{j}},u_{k}^{\epsilon_{k}})$ leave
the $\epsilon_{j}$ side of vertex $u_{j}$ and arrive at the $\epsilon_{j}$
side of $u_{j}$. See Figure \ref{rules} for the diagrams representing $O(\theta_1,\theta_2,u^+,u^-)$. 
The connected part of the multiparticle octagon
consists of exactly those graphs which are connected. Actually the sum of these
terms are not singular and takes a very special finite form. For one pair of mirror particles
we have 
\begin{eqnarray}
O^c(\theta_{1},\theta_{2},u^{+},u^{-})&=& O(\theta_{1},u^{+})O(\theta_{2},u^{-})+O(\theta_{1},u^{-})O(\theta_{2},u^{+})  \nonumber \\
&=&O(\theta_{1},\theta_{2})(k(u-\theta_{1})+k(u-\theta_{2})) \quad ;\qquad k(u)=-\frac{1}{\cosh u}
\end{eqnarray}
 which generalize to $n$ particles as 
\begin{equation}
O^c(\theta_{1},\theta_{2},u_{1}^{+},u_{1}^{-},\dots,u_{n}^{+},u_{n}^{-})=
O(\theta_{1},\theta_{2})\prod_{i=1}^n(k(u_{i}-\theta_{1})+k(u_{i}-\theta_{2}))
\end{equation}
This can be checked by noticing that the connected form factors satisfy the kinematical
singularity axiom at $\theta_i+i\pi/2=u_j $. Resolving the related recursion 
leads to the formula of the connected form factors.

Taking naively the measures to be trivial $\mu(\{u_{i}\})=1$ leads
to the naive and ``wrong'' result
\begin{equation}
N_{L}(\theta_{1},\theta_{2})=O(\theta_{1},\theta_{2})d_{\mathrm{n}}(\theta_{1})d_{\mathrm{n}}(\theta_{2})\quad;\qquad d_{\mathrm{n}}(\theta)=e^{\int_{-\infty}^{\infty}\f{du}{2\pi}k(u-\theta)e^{-mL\cosh u}}
\end{equation}
At the leading wrapping order the naive result is correct, indicating
that we are missing some relevant contributions from multiple wrappings.
Actually the missing terms come from the domains of integrations,
when $u_{i}=u_{j}$ for some $i$ and $j$. As a guiding principle
one could demand that $N_{L}(\theta_{1},\theta_{2})$ should satisfy
the DSFT vertex axioms. In Appendix B we show that it is equivalent
to the teleportation requirement, which can be rephrased as that after
an analytical continuation $\theta\to\theta-i\pi$ we have 
\begin{equation}
d_{L}(\theta+i\pi)=(1-e^{-ipL})\frac{1}{d_{L}(\theta)}
\end{equation}
Expanding both sides and taking into account that $k(\theta-i\pi)=-k(\theta)$
we can see that the residue terms must sum up to $-e^{-ipL}/d_{L}(\theta)$.
Evaluating at leading order gives 
\begin{equation}
\mu_{1}(u)=1+O(e^{-mL\cosh u})
\end{equation}
At next order, assuming $ \mu_{2}(u_1,u_2)=1+O(e^{-mL\cosh u})$, the direct continuation of $d_{L}(\theta)$ from the double pole term produces
a term $\frac{1}{2}e^{-2ipL}$, which can be canceled choosing 
\begin{equation}
\mu_{1}(u)=1+\frac{1}{2}e^{-mL\cosh u}+O(e^{-2mL\cosh u})
\end{equation}
Calculating systematically the higher order terms we can find that
\begin{equation}
\mu_{1}(u)=\sum_{n=0}^\infty\frac{1}{n+1}e^{-nmL\cosh u}
\qquad;\qquad\mu_{n}(\{u_{i}\})=\prod_{i}\mu_{1}(u_{i})
\end{equation}
which gives the expected results: 
\begin{equation}
\mu_{1}(u)e^{-mL\cosh u}=\log\left(1-e^{-m L \cosh u}\right)
\end{equation}
 Clearly the relevant nontrivial terms are kinds
of ``diagonal'' contribution associated with multiple wrapping. 
In order to understand better their role and origin we will now look at the
gluing process from the point of view of so-called cluster expansion in relativistic
integrable field theories. Then we will revisit again the gluing of the octagon as
well as describe how one can glue the decompactified SFT vertex into the finite volume one.

\section{The structure of multiple wrapping corrections}

In this section we first exhibit explicitly the exactly known observables
for a free massive boson which will give insight to the multiple wrappings,
starting from the completely standard examples of free energy and
going on to the quite intricate formulas for the exact string vertex.
We then analyze the general structure of the wrapping corrections.

\subsubsection*{The ground state energy}

The ground state energy can be obtained from the large $R$ limit
of the torus partition function 
\begin{equation}
Z\sim e^{-RE_{0}(L)}
\end{equation}
with 
\begin{equation}
E_{0}(L)=\pm m\int_{-\infty}^{\infty}\f{d\th}{2\pi}\cosh\th\log\left(1\mp e^{-mL\cosh\th}\right)
\end{equation}
with the upper/lower signs corresponding to a free boson/fermion.
Expanding the above formula in a power series in $e^{-mL\cosh\th}$
gives multiple wrapping contributions to the ground state energy.

Incidentally the exact equation which holds in the interacting case
has a very similar form 
\begin{equation}
E_{0}(L)=-m\int_{-\infty}^{\infty}\f{d\th}{2\pi}\cosh\th\log\left(1+e^{-\eps(\th)}\right)
\end{equation}
where $\eps(\th)$ is a solution of the relevant TBA equation.

\subsubsection*{The LeClair-Mussardo formula}

The finite volume expectation value of a local operator is given by
the following formula \cite{Leclair:1999ys}
\begin{equation}
\cor{\OO}_{L}=\sum_{n=0}^{\infty}\int\prod_{k=1}^{n}\f{d\th_{k}}{2\pi}
\f{1}{1\mp e^{mL\cosh\th_{k}}}F_{n}^{c}(\th_{1},\ldots,\th_{n})
\end{equation}
Here $F_{n}^{c}(\th_{1},\ldots,\th_{n})$ is the infinite volume (connected)
diagonal form factor of the operator $\OO$. Remarkably enough the
above formula again generalizes to the interacting case through the
simple substitution $mL\cosh\th\to\eps(\th)$ \cite{Leclair:1999ys}: 
\begin{equation}
\cor{\OO}_{L}=\sum_{n=0}^{\infty}\int\prod_{k=1}^{n}\f{d\th_{k}}{2\pi}
\f{1}{1+e^{\eps(\th_{k})}}F_{n}^{c}(\th_{1},\ldots,\th_{n})
\end{equation}

For completeness let us quote here the finite volume expansions for the
decompactified SFT vertex as well as the finite size string vertex.
It is illuminating to recognize the structural similarity of the 
multiple wrapping terms appearing in these expressions with the
relativistic formulas given above.

\subsubsection*{The decompactified string vertex}

The formula for the decompactified string vertex with two particles
on string \#2 takes the form:
\begin{equation}
N_{L}(\th_{1},\th_{2})=O(\th_{1},\th_{2})d_{L}(\th_{1})d_{L}(\th_{2})
\end{equation}
where the logarithm of the function $d_{L}(\th)$ is 
\begin{equation}
\log d_{L}(\th)=-\int_{-\infty}^{\infty}\frac{du}{2\pi}\,k(u-\theta)\log(1-e^{-mL\cosh u})
\end{equation}
We note a surprising similarity with the ground state energy formula.

\subsubsection*{The finite size string vertex}

The formula for the string vertex with all the three strings being
of finite size $L_{i}$ has been derived by a direct calculation in
\cite{LSNS}. The formulas there can be recast into a simpler form
when expressed in terms of rapidities and take the form (again up to an overall normalization):
\begin{equation}
N_{L_{1}}^{L_{2}}(\th_{1},\th_{2})=N_{L_{1}}(\th_{1},\th_{2})
\cdot\f{d_{L_{2}}(\th_{1})}{d_{L_{3}}(\th_{1})}\cdot\f{d_{L_{2}}(\th_{2})}{d_{L_{3}}(\th_{2})}
\end{equation}

\subsection{Mirror channel compactification -- cluster expansion}

Let us now review the approach of mirror channel compactification {\it aka} cluster expansion.
Let us first consider the case of the partition function evaluated on a torus of size $L\times R$
where $R$ is very large in order to extract the ground state.
As in the derivation of the TBA, it is convenient to perform this calculation in the mirror
channel (which is compactified to the large size $R$). Then the partition function is by definition
the summation over all states in the mirror theory weighted with $e^{-EL}$ where $E$ is the mirror channel
energy:
\begin{equation}
\label{e.part}
1+\sum_n e^{-E_n L}+ \sum_{n_1\geq n_2} e^{-(E_{n_1}+E_{n_2}) L} +\ldots
\end{equation}
Here for the free boson, mode numbers can coincide hence we have $n_1\geq n_2$. In contrast for an interacting
theory or for a free fermion we would have a sharp inequality.
In the next step one takes the continuum limit
\eq
\sum_n \rightarrow R\int_{-\infty}^\infty \f{d\th}{2\pi} m \cosh \th
\eqx
however, and this is the key point, one has to take care of the diagonal terms and first
separate them out
\eq
\label{e.discrete}
\sum_{n_1\geq n_2}=\f{1}{2} \sum_{n_1,n_2} + \f{1}{2} \sum_{n_1=n_2}
\eqx
Writing all the contributions appearing in (\ref{e.part}) gives
\eqn
&&1+R\int_{-\infty}^\infty \f{d\th}{2\pi} m \cosh \th \,e^{-mL\cosh\th} +
\f{1}{2} R\int_{-\infty}^\infty \f{d\th}{2\pi} m \cosh \th \,e^{-2mL\cosh\th}+ \nonumber\\
&& + \f{1}{2} \left( R\int_{-\infty}^\infty \f{d\th}{2\pi} m \cosh \th \,e^{-mL\cosh\th} \right)^2 +\ldots
\eqnx
We see that this coincides with the first terms of the expansion of
\eq
e^{-m R \int_{-\infty}^{\infty}\f{d\th}{2\pi}\cosh\th\log\left(1- e^{-mL\cosh\th}\right)}
\eqx
The diagonal terms like the one with $n_1=n_2$ are exactly responsible for the nontrivial measure factor 
$\log\left(1- e^{-mL\cosh\th}\right)$.
Let us emphasize that their interpretation is not as straightforward as it may seem.
Indeed, for an interacting theory such states with $n_1=n_2$ are \emph{not} even part of the spectrum.
Provisionally a useful interpretation of these terms is that they represent \emph{multiply wrapped}
single particles. This interpretation seems to provide the correct intuition for the treatment of such terms
in all cases considered in this paper.

The LeClair-Mussardo formula arises when we insert a local operator into the above expansion.
We thus have to evaluate the diagonal expectation values of the type
\eq
\corr{n_{1}n_{2}}{\OO}{n_{2}n_{1}}_{R}
\eqx
for asymptotically large $R$, i.e. neglecting any wrapping terms in $R$.
In this limit the expectation value can be written\footnote{This formula is still
conjectural for more than two particles but there is overwhelming evidence that it is correct \cite{Pozsgay:2007gx,Pozsgay:2010xd}.} 
as a linear combination of appropriate measure factors (with the only explicit $R$ dependence) and
\emph{infinite volume} diagonal form factors of the local operator $\OO$.
For the case of a free massive boson we have the following explicit formulas for up to two particles
\eqn
\corr{n_1}{\OO}{n_1}_R &=& \f{F_1^c(\th_1)}{R E_1}+ F^c_0 \\
\corr{n_1 n_2}{\OO}{n_2 n_1}_R  &=&   \f{F_2^c(\th_1,\th_2)}{R^2 E_1E_2}+ 
 \f{F_1^c(\th_1)}{R E_1}+
\f{F_1^c(\th_2)}{R E_2}+ F^c_0 \\
\corr{{n_1}{n_1}}{\OO}{{n_1}{n_1}}_R &=& 0 + {\mathbf 2} \cdot \f{F_1^c(\th_1)}{R E_1}+ F^c_0
\eqnx
Note the factor of $2$ in the diagonal double wrapping term. It is exactly this factor (and other factors of
this type at higher orders) that cancels the $1/2$ appearing in front of the diagonal term in (\ref{e.discrete})
and effectively transforms the measure factor
\eq
\log\left(1- e^{-mL\cosh\th}\right)
\eqx
into
\eq
\f{-1}{1- e^{mL\cosh\th}}
\eqx
appearing in the LeClair-Mussardo formula.

\subsection{The structure of the multiple wrappings }

Looking at the above two examples, we see that the computation of
the finite volume observable $\cor{\XX}_{L}$ can be summarized as
regularizing the mirror channel (e.g. by compactifying it on a finite
but large volume $R$), decomposing the summation over a complete
basis of states into independent sums of single and multiple wrapped
particles with appropriate combinatorial factors, and finally providing
an expression for the \emph{diagonal} finite volume \emph{asymptotic}\footnote{i.e. neglecting all $e^{-mR}$ terms}
expectation values, namely 
\begin{eqnarray}
Z\cor{\XX}_{L} & = & \corr{\vac}{\XX}{\vac}_{R}+\sum_{n_{1}}\corr{n_{1}}{\XX}{n_{1}}_{R}e^{-E_{n_{1}}L}+\f{1}{2}\sum_{n_{1},n_{2}}\corr{n_{1}n_{2}}{\XX}{n_{2}n_{1}}_{R}e^{-(E_{n_{1}}+E_{n_{2}})L}\label{e.clustergen}\\
 &  & \hspace{-1.5cm}+\f{1}{2}\sum_{n_{1}}\corr{{n_{1}}^{(\times2)}}{\XX}{{n_{1}}^{(\times2)}}_{R}e^{-2E_{n_{1}}L}+\f{1}{6}\sum_{n_{1},n_{2},n_{3}}\corr{n_{1}n_{2}n_{3}}{\XX}{n_{3}n_{2}n_{1}}_{R}e^{-(E_{n_{1}}+E_{n_{2}}+E_{n_{3}})L}\nonumber \\
 &  & \hspace{-1.5cm}+\f{1}{2}\sum_{n_{1},n_{2}}\corr{n_{1}n_{2}^{(\times2)}}{\XX}{n_{2}^{(\times2)}n_{1}}_{R}e^{-(E_{n_{1}}+2E_{n_{2}})L}+\f{1}{3}\sum_{n_{1}}\corr{{n_{1}}^{(\times3)}}{\XX}{{n_{1}}^{(\times3)}}_{R}e^{-3E_{n_{1}}L}+\ldots\nonumber 
\end{eqnarray}
In an interacting or fermionic theory one has to flip some signs as
there all quantized mode numbers must be distinct. The key remaining
information are the above $R$-regularized diagonal expectation values.
We may expect that they have the following general form 
\begin{equation}
\label{e.fvevgen}
\corr{\{n_{i}^{(\times k_{i})}\}}{\XX}{\{n_{i}^{(\times k_{i})}\}}_{R}=
\sum_{\al\cup\alb}\mu(\al,\alb,R)\cdot\FF_{\XX}(\{n_{i}^{(\times k_{i})}\}_{i\in\al})
\end{equation}
The measure factor is the only place with explicit $R$ dependence.
For the free boson we expect it to take a simple form 
\begin{equation}
\mu(\al,\alb,R)=\f{1}{\prod_{i\in\al}RE_{i}}
\end{equation}
The second factor in (\ref{e.fvevgen}) should be a quantity defined in infinite volume
associated to the observable $\XX$ which should follow from some
appropriate functional equations. 

Alternatively we could calculate $\cor{\XX}_{L}$ instead of $Z\cor{\XX}_{L}$.
By this we remove many disconnected terms (as $Z^{-1}$ has the same
structure as $\cor{\XX}_{L}$.) The modified quantity appearing in
the expansion is denoted by $\FF_{\XX}^{c}(\{n_{i}^{(\times k_{i})}\}_{i\in\al})$
where the superscript $^{c}$ indicates that one would have to take
just the connected part. Note that care should be taken to define
these generalized form factors also for the multiply wrapped particles.
How to do it in general is by no means obvious. The main result of
this paper is to provide the relevant expressions both for the octagon
and for the decompactified string vertex with two external particles
such that summing (\ref{e.clustergen}) for the octagon yields the
decompactified string vertex and summing (\ref{e.clustergen}) for
the decompactified string vertex gives the exact finite volume string
vertex.

Before doing so, we summarize the relevant quantities both for the
ground state energy and for the LeClair-Mussardo formula.

\subsubsection*{Ground state energy}

The ground state energy is related to the torus partition function as $Z\sim e^{-RE_{0}(L)}$,
thus we basically analyze the $\XX=\mathbb{I}$ situation. In this
case the $\alpha\cup\bar{\alpha}$ decomposition degenerates only
to one ``trivial'' term 
\begin{equation}
\corr{\{n_{i}^{(\times k_{i})}\}}{\XX}{\{n_{i}^{(\times k_{i})}\}}_{R}=1=
\frac{1}{\prod_{i}RE_{i}}\FF_{\XX}(\{n_{i}^{(\times k_{i})}\})
\end{equation}
In particular, in our normalization it implies that 
\begin{equation}
\FF_{\XX}(\{n_{i}^{(\times k_{i})}\})=\prod_{i}RE_{i}
\end{equation}

\subsubsection*{LeClair-Mussardo formula}

In the case of the LeClair-Mussardo formula we analyze the expansion
of $\cor{\OO}_{L}$ instead of $Z\cor{\OO}_{L}$, since by this trick
we can remove all disconnected terms and the $\alpha\cup\bar{\alpha}$
decomposition degenerates only to one term 
\begin{equation}
\FF_{\XX}^{c}(\{n_{i}^{(\times k_{i})}\})=F_{n}^{c}(\{n_{i}\})\prod_{i}k_{i}
\end{equation}
where $F_{n}^{c}(\{n_{i}\})$ is the connected diagonal form factor. Thus the wrapping order appears only as a combinatorial factor.

\section{Resumming the octagon}

Here we normalize the decompactified string vertex as $N_{L}(\emptyset)=1$,
i.e. we calculate only the connected contributions (see section~\ref{s.naiveoctagon} for
their definition). We propose the
following form of the finite volume expectation values which, when
inserted into (\ref{e.clustergen}) will exactly reproduce the decompactified
string vertex with string \#1 being of length $L$: 
\begin{eqnarray}
\corr{\vac}{\oct}{\vac}_{R} & = & O(\th_{1},\th_{2})\nonumber \\
\corr{n_{1}}{\oct}{n_{1}}_{R} & = & \f{1}{RE_{1}}O^{c}(\th_{1},\th_{2},u_{1}^{-},u_{1}^{+})\equiv\f{1}{RE_{1}}O(\th_{1},\th_{2})\left(k(u_{1}-\th_{1})+k(u_{1}-\th_{2})\right)\nonumber \\
\corr{n_{1}n_{2}}{\oct}{n_{2}n_{1}}_{R} & = & \f{1}{R^{2}E_{1}E_{2}}O^{c}(\th_{1},\th_{2},u_{1}^{-},u_{2}^{-},u_{2}^{+},u_{1}^{+})\nonumber \\
 & = & \f{1}{R^{2}E_{1}E_{2}}O(\th_{1},\th_{2})\prod_{i=1}^{2}\left(k(u_{i}-\th_{1})+k(u_{i}-\th_{2})\right)
\end{eqnarray}
The key assumption now involves the expectation values when some of
the mirror particles wrap multiple times. The exact answer implies
that the relevant expectation value does not depend on the wrapping
order. E.g. we have 
\begin{equation}
\corr{{n_{1}}^{(\times2)}}{\oct}{{n_{1}}^{(\times2)}}_{R}=\corr{n_{1}}{\oct}{n_{1}}_{R}=\f{1}{RE_{1}}O(\th_{1},\th_{2})\left(k(u_{1}-\th_{1})+k(u_{1}-\th_{2})\right)
\end{equation}
With the above formulas in place, it is simple to generalize as
\begin{eqnarray}
\corr{\{{n_{i}}^{(\times k_{i})}\}}{\oct}{\{{n_{i}}^{(\times k_{i})}\}}_{R}
&=&\corr{\{n_{i}\}}{\oct}{\{n_{i}\}}_{R}\nonumber \\ &=&O(\th_{1},\th_{2})\prod_{i}\f{1}{RE_{i}}\left(k(u_{i}-\th_{1})+k(u_{i}-\th_{2})\right)
\end{eqnarray}
As then one can easily convince oneself (looking e.g. at all terms
up to $3^{rd}$ order or comparing to the free boson ground state energy) 
that (\ref{e.clustergen}) can be summed up
to give exactly the decompactified string vertex 
\begin{eqnarray}
N_{L}(\th_{1},\th_{2}) & = & O(\th_{1},\th_{2})\cdot
\underbrace{e^{-\int_{-\infty}^{\infty}\f{du}{2\pi}k(u-\theta_{1})
\log(1-e^{-mL\cosh u})}}_{d_{L}(\th_{1})}\cdot(\th_{1}\to\th_{2})\nonumber \\
 & = & O(\th_{1},\th_{2})d_{L}(\th_{1})d_{L}(\th_{2})
\end{eqnarray}

\section{Resumming the string vertex}

Passing from the decompactified string vertex with string \#1 being
of size $L_{1}$ and strings \#2 and \#3 being infinite to the finite
volume string vertex with all strings having finite size $L_{i}$
can be done following closely the strategy employed for the octagon.
We will again consider a configuration with two particles on string
\#2. In order to do the infinite volume part of the calculation it
is convenient to transport the mirror particles up to string \#2.
Now the rapidities will have to be shifted by $\pm i\pi/2$ so in
this section we will denote 
\begin{equation}
u^{\pm}=u\pm i\f{\pi}{2}
\end{equation}
Since we are dealing with a free theory the decompactified vertex
with multiple particles will be obtained by Wick contractions but
now with pairing performed with 
\begin{equation}
N_{L}^{\infty}(\th_{1},\th_{2})=N_{L_{1}}(\th_{1},\th_{2})=O(\th_{1},\th_{2})d_{L_{1}}(\th_{1})d_{L_{1}}(\th_{2})
\end{equation}
So we see that the nontrivial part of the computation is almost exactly
the same as for the octagon (up to the redefinition of $u^{\pm}$
here) and the $L_{1}$-dependent factors will appear only as an overall
product for all particles entering the amplitude. It is clear that
we thus get the following expressions: 
\begin{align}
 & \corr{\vac}{\neum}{\vac}_{R}=N_{L_{1}}^{\infty}(\th_{1},\th_{2})\\
 & \corr{n_{1}}{\neum}{n_{1}}_{R}=\f{1}{RE_{1}}N_{L_{1}}^{\infty}(\th_{1},\th_{2})\left(k(u_{1}-\th_{1})+k(u_{1}-\th_{2})\right)d_{L_{1}}(u_{1}^{+})d_{L_{1}}(u_{1}^{-})\label{e.neumsingle}\\
 & \corr{n_{1}n_{2}}{\neum}{n_{2}n_{1}}_{R}=\f{1}{R^{2}E_{1}E_{2}}N_{L_{1}}^{\infty}(\th_{1},\th_{2})\prod_{i=1}^{2}\left(k(u_{i}-\th_{1})+k(u_{i}-\th_{2})\right)d_{L_{1}}(u_{i}^{+})d_{L_{1}}(u_{i}^{-})\label{e.neumtwo}
\end{align}
The product of the $d_{L_{1}}(.)$ functions can be simplified using
the functional equations 
\begin{equation}
d_{L_{1}}(u^{+})d_{L_{1}}(u^{-})=\left(1-e^{-mL_{1}\cosh u}\right)
\end{equation}
In order to see the crucial role of the above remaining $u$-dependent factor let us consider
the expression (\ref{e.clustergen}) up to the single particle term.
We have 
\begin{equation}
N_{L_{1}}^{\infty}(\th_{1},\th_{2})\left[1+\int_{-\infty}^{\infty}\f{du}{2\pi}\left(k(u-\th_{1})+k(u-\th_{2})\right)
e^{-mL_{2}\cosh u}\left(1-e^{-mL_{1}\cosh u}\right)+\ldots\right]
\end{equation}
The 1 particle term thus splits into a difference of two terms: one
with a wrapping factor $e^{-mL_{2}\cosh u}$ and the other with the
wrapping factor $e^{-m(L_{1}+L_{2})\cosh u}\equiv e^{-mL_{3}\cosh u}$,
where we used the conservation of lengths in the light cone gauge
$L_{1}+L_{2}=L_{3}$. But these are indeed exactly the first terms
in the expansion of 
\begin{equation}
\f{d_{L_{2}}(\th_{1})}{d_{L_{3}}(\th_{1})}\cdot\f{d_{L_{2}}(\th_{2})}{d_{L_{3}}(\th_{2})}\label{ds}
\end{equation}
It is clear that the two particle term coming from (\ref{e.neumtwo})
will contribute to the exponentiation of the above structure. However
the contribution of the doubly wrapped particle is quite subtle and
requires some care. In order to motivate our proposal, let us recall
that the decompactified string vertex axioms introduced in \cite{SFT}
involve an overall monodromy factor $e^{ipL_{1}}$. Now since we are
considering a particle which wraps twice across the vertex, we expect
that it would effectively feel a factor $e^{2ipL_{1}}$. Thus it is
very natural to expect that the generalization of formula (\ref{e.neumsingle})
to a doubly wrapped particle takes the form 
\begin{equation}
\corr{{n_{1}}^{(\times2)}}{\neum}{{n_{1}}^{(\times2)}}_{R}=
\f{1}{RE_{1}}N_{L_{1}}^{\infty}(\th_{1},\th_{2})\left(k(u_{1}-\th_{1})
+k(u_{1}-\th_{2})\right)d_{2L_{1}}(u_{1}^{+})d_{2L_{1}}(u_{1}^{-})
\end{equation}
Now we can examine the doubly wrapped particle contribution
in (\ref{e.clustergen}): 
\begin{equation}
N_{L_{1}}^{\infty}(\th_{1},\th_{2})\cdot\f{1}{2}\int_{-\infty}^{\infty}\f{du}{2\pi}\left(k(u-\th_{1})+k(u-\th_{2})\right)e^{-2mL_{2}\cosh u}\left(1-e^{-2mL_{1}\cosh u}\right)
\end{equation}
We see that this yields the first nontrivial double wrapping terms
in the expansion of the logarithms in (\ref{ds}). In order for this
to work it was absolutely crucial that the double wrapped particle
feels effectively the double factor $e^{2ipL_{1}}$. It is clear that
analogous property should hold for multiple wrapped particles.
\begin{equation}
\corr{\{{n_{i}}^{(\times k_{i})}\}}{\neum}{\{{n_{i}}^{(\times k_{i})}\}}_{R}=
N_{L_{1}}^{\infty}(\th_{1},\th_{2})\prod_{i}\f{d_{k_{i}L_{1}}(u_{i}^{+})d_{k_{i}L_{1}}(u_{i}^{-})}{RE_{i}}\left(k(u_{i}-\th_{1})+k(u_{i}-\th_{2})\right)
\end{equation}

Repeating the above for higher number of particles we see that we
obtain the exact finite volume Neumann coefficient 
\begin{equation}
N_{L_{1}}^{L_{2}}(\th_{1},\th_{2})=N_{L_{1}}(\th_{1},\th_{2})\cdot\f{d_{L_{2}}(\th_{1})}{d_{L_{3}}(\th_{1})}\cdot\f{d_{L_{2}}(\th_{2})}{d_{L_{3}}(\th_{2})}\equiv O(\th_{1},\th_{2})\cdot\f{d_{L_{1}}(\th_{1})d_{L_{2}}(\th_{1})}{d_{L_{3}}(\th_{1})}\cdot\f{d_{L_{1}}(\th_{2})d_{L_{2}}(\th_{2})}{d_{L_{3}}(\th_{2})}
\end{equation}

\section{Conclusions}

The nontrivial topology of the SFT vertex allows for various lines of approach towards
determining it exactly. By cutting the vertical edges various number of times and decompactifying
one obtains the decompactified string vertex of \cite{SFT}, the octagon and two hexagons of \cite{HEXAGON}.
Although the final goal is the determination of the exact finite volume vertex, i.e. with all the three
strings being of finite size, the necessity of passing through this intermediate decompactified stage
is that only then we can formulate functional equations for the relevant quantities which 
incorporate analyticity and various variations of crossing symmetry.
One of the contributions of the present paper was to formulate appropriate axioms for the octagon
in the interacting case.
 
Hence a key question is to understand the procedure of gluing back the decompactified answers 
into the final finite volume result.
In \cite{HEXAGON} a formal expression for gluing back was suggested by a summation over 
a complete set of mirror particles living on the edge which is being glued.
This expression is, however, rather formal as it stands and suffers from divergences.
The subtleties arise at the multiple wrapping level which is in general difficult to study.

The case of the pp-wave vertex (essentially a free massive boson on the string pants diagram)
is a very interesting theoretical laboratory for studying these issues as we have at our disposal
exact finite volume answers for the finite size SFT vertex as well as its various decompactified variations ---
the decompactified SFT vertex and the octagon. As these expressions are exact and incorporate
an infinite set of multiple wrapping corrections we may quantitatively explore the subtleties
of the gluing procedure.

We argue that the quantitative structure of the gluing procedure may be efficiently understood
within the so-called cluster expansion (equivalently compactification in the mirror channel).
There the main ingredient is the asymptotic large mirror volume expectation value for the observable
in question which should decompose into a linear combination of measure factors and appropriate
infinite volume quantities. This is a standard way to understand ground state energy and the LeClair-Mussardo
formula for one point expectation values in relativistic integrable theories.
In the present paper we adopt this framework to the case of the octagon and the decompactified SFT vertex.
Note, however, that even in the classical case of LeClair and Mussardo there is no proof of the
general large mirror volume expectation value formula for more than two particles.
In the case of the vertex we also do not provide a proof, however our proposed formulas
are very natural from the physical point of view. Also \emph{a-posteriori} it is very
nontrivial that any such formulas exist which reproduce the apparently very complicated 
finite volume Neumann coefficients.

We demonstrated that one can resum the multiple wrapping corrections for the octagon
into the exact decompactified SFT vertex. This necessitates a nontrivial, but quite natural modification
of the multiple wrapping measure. We then proceed to interpret this modification through the
cluster expansion where it turns out to arise from certain diagonal terms. We then show that similarly one can
resum the decompactified SFT vertex and recover the exact finite volume pp-wave Neumann coefficients.

There are numerous further questions to investigate. A key question, and one of the long term motivations
of this work, would be to guess some underlying exact TBA-like formulation for the SFT vertex.
The integral expression for the pp-wave Neumann coefficient obtained in the present paper
is very intriguing in that respect. Also in this paper we did not discuss the hexagons at all.
It would be interesting to understand this better, as well as the differences 
w.r.t. \cite{HEXAGON}\footnote{A main difference between the approach of \cite{HEXAGON} and the
considerations of the present paper is that there the light cone gauge choice is different
for each of the three strings, while here we concentrate on the conventional light cone SFT vertex
picture where we have a single gauge choice, and hence e.g. the total size of the strings is conserved.}.

In this paper we focused on the 3-point functions and on the way how they could be 
described by gluing octagons and the DSFT vertex. The 4-point 
functions, however, are even more interesting and recently there 
have been activities using integrable methods in their descriptions
\cite{Basso:2017khq,Bargheer:2017eoz,Fleury:2016ykk,Eden:2016xvg}
It would be very challenging to figure out how two octagons (or 
their modifications) could be glued together to describe the 
four point functions. Actually the geometry of the 4-point function
allows for two different cuttings into two octagons. Demanding 
their compatibility might lead to non-trivial constraint on the 
gluing procedure or on the octagon themselves. 

\bigskip

{\bf Acknowledgments.} RJ was supported by NCN grant 2012/06/A/ST2/00396 and ZB by a Lend\"ulet and by the  NKFIH 116505 Grant. RJ would like to thank the Galileo Galilei Institute for Theoretical Physics for
hospitality and the INFN for partial support during the completion of this work.

\appendix

\section{Large volume expansion of the plane-wave DSFT vertex}

In this Appendix we rewrite the plane-wave DSFT vertex into the form,
in which multiple wrapping terms can be easily identified. Recall from \cite{LSNS,SFT}
that the DSFT vertex takes the form\footnote{Here we normalized the DSFT vertex as 
$N_L(\emptyset)=1$.} 
\begin{equation}
N(\theta_{1},\theta_{2})=O(\theta_{1},\theta_{2})d_{L}(\theta_{1})d_{L}(\theta_{2})
\end{equation}
where 
\begin{equation}
d_{L}(\theta)=\pi \sqrt{\frac{2}{ML}}\frac{e^{\frac{\theta}{2\pi}pL}}{\sinh\frac{\theta}{2}}\frac{1}{\tilde{\Gamma}_{\frac{mL}{2\pi}}(\theta)}
\end{equation}
We are interested in the large $L$ expansion of $\tilde{\Gamma}_{\frac{mL}{2\pi}}$. The large $L$ expansion of 
$\Gamma_\mu (z)$ was calculated in \cite{LSNS} and rephrased in \cite{SFT} as 
\begin{equation}
\tilde{\Gamma}_{\mu}(\theta)=\sqrt{\frac{\pi}{\mu}}
\frac{e^{\frac{\theta}{2\pi}pL}}{\sinh\frac{\theta}{2}}\tilde{\Gamma}_{\exp}(\theta) \quad ;\qquad
\mu =\frac{ML}{2\pi}
\end{equation}
where $\tilde{\Gamma}_{\mathrm{exp}}(\theta)$ vanishes exponentially
for large $L$ in the following way: 
\begin{equation}
\partial_{\mu}\log\tilde{\Gamma}_{\exp}(z)=-2\sum_{n=1}^{\infty}\frac{\mu}{\omega(z)}K_{0}(2\mu n\pi)=-\frac{\mu}{\omega(z)}\int_{-\infty}^{\infty}du\frac{1}{e^{mL\cosh u}-1}
\end{equation}
We can integrate this equation as 
\begin{eqnarray}
\log\tilde{\Gamma}_{\exp}(z) & = & -\int_{-\infty}^{\infty}\frac{du}{2\pi}\,\frac{\log(1-e^{-2\pi\mu\cosh u})}{\sqrt{1+\frac{z^{2}}{\mu^{2}}}\cosh u-\frac{z}{\mu}\sinh u}
\end{eqnarray}
and the constant of integration is fixed from the vanishing large
volume limit. By introducing the rapidity variable via $z=\mu\cosh\theta$
we get 
\begin{equation}
\tilde{\Gamma}_{\exp}(\theta)=\exp\left\{ \int_{-\infty}^{\infty}\frac{du}{2\pi}\,k(u-\theta)\log(1-e^{-mL\cosh u})\right\} \quad;\qquad k(\theta)=-\frac{1}{\cosh(u-\theta)}
\end{equation}
This implies that 
\begin{equation}
\log d_{L}(\theta)=-\log\tilde{\Gamma}_{\exp}(\theta)=\sum_{n=1}^{\infty}\frac{1}{n}\int_{-\infty}^{\infty}\frac{du}{2\pi}\,k(u-\theta)e^{-nmL\cosh u}
\end{equation}
In order to check this expression we first perform an analytical continuation
in $\theta$ as $\theta\to\theta+i\pi$. In doing so a pole singularity
of the kernel $k(u-\theta)$ crosses the integration contour, which
contribute to the functional relation giving
\begin{equation}
d_{L}(\theta+i\pi)=(1-e^{-ipL})\frac{1}{d_{L}(\theta)}
\end{equation}
which is required by the kinematical singularity axiom. 
Continuing further to $\theta\to\theta+2i\pi$ another singularity
crosses the integration contour, which contributes with an opposite
residue leading to 
\begin{equation}
d_{L}(\theta+2i\pi)=\frac{1-e^{ipL}}{1-e^{-ipL}}d_{L}(\theta)=-e^{ipL}d_{L}(\theta)
\end{equation}
which is the required monodromy property of the function.

\section{DSFT vertex axioms from octagon axioms}

In this Appendix we show how the DSFT vertex axioms could be obtained
from the octagon axioms.  This will shed also light,
how we need to use the octagon amplitude to describe particles in
the split \#3 and \#3' domains. For simplicity we present the ideas 
for the free theory and for 2 particles only. The generalization for the interacting theory can be easily 
done at the formal level similarly to eq. (\ref{formal}). At a less formal level one has to understand how to regularize 
the kinematical singularities for the mirror particle-anti-particle pairs.

Recall that the DSFT vertex 
can be written in terms of the connected octagons as
\begin{eqnarray}
N_{L}(\theta_{1},\theta_{2}) & = & O(\theta_{1},\theta_{2})+\int_{-\infty}^{\infty}\frac{du}{2\pi}\mu_{1}(u)O^c(\theta_{1},\theta_{2},u^+,u^-)e^{-LE(u)}+\\
 &  & \frac{1}{2}\int_{-\infty}^{\infty}\frac{du_{1}}{2\pi}\int_{-\infty}^{\infty}\frac{du_{2}}{2\pi}\,\mu_{2}(u_{1},u_{2})O^c(\theta_{1},\theta_{2},u_{1}^+,u_{2}^+,u_{2}^-,u_{1}^-)e^{-L\left(E(u_{1})+E(u_{2})\right)}+
 \dots \nonumber  
\end{eqnarray}
where $u^{\pm}=u\pm \frac{3i\pi}{2}$. Let us see now how the DSFT axioms are satisfied.
\begin{itemize}
\item The exchange axiom is trivially reproduced as each term in the expansion
has this property. In the free boson theory the connected and the
disconnected terms are mapped to each other under the exchange $\theta_{1}\leftrightarrow\theta_{2}$,
thus the connected terms are symmetric themselves. 
\item In order to show the kinematical singularity axiom we continue analytically
$\theta_{1}\to\theta_{1}+i\pi$. As a first step we continue it into
the mirror domain between space \#2 and \#3: $\theta_{1}\to\theta_{1}+\frac{i\pi}{2}$.
When it is exactly in the mirror domain it will hit a kinematical
singularity of the octagon coming from integrals for mirror particles
of type $u_{i}^{+}$. We can avoid this singularity by slightly deforming
the contours. However, when we continue the particle's rapidity further
to domain \#3 we cross the integration contour by a pole singularity.
Thus we will have two types of contributions: the direct continuations and the
pole contributions. See Figure \ref{cont1} for a graphical representation. 
The direct term, denoted by $N_{\mathrm{sum}}(\theta_{1}+i\pi ,\theta_{2})$ is simply 
\begin{eqnarray}
 &  & O(\theta_{1}+i\pi,\theta_{2})+\int_{-\infty}^{\infty}\frac{du}{2\pi}\mu_{1}(u)O^c(\theta_{1}+i\pi,\theta_{2},u^+,u^-)e^{-LE(u)}+\\
 &  & \frac{1}{2}\int_{-\infty}^{\infty}\frac{du_{1}}{2\pi}\int_{-\infty}^{\infty}\frac{du_{2}}{2\pi}\,\mu_{2}(u_{1},u_{2})
O^c(\theta_{1}+i\pi,\theta_{2},u_{1}^+,u_{2}^+,u_{2}^-,u_{1}^-)e^{-L\left(E(u_{1})+E(u_{2})\right)}+\dots \nonumber
\label{Osum}  
\end{eqnarray}
Let us explain the notation a bit. In the following we understand by $N_{\mathrm{sum}}(\theta_{1}+i\pi ,\theta_{2})$ the above sum. 
So whatever is the argument of  $N_{\mathrm{sum}}(\theta_1,\theta_2)$ it means we evaluate the octagon sum at that rapidities and we 
\emph{do not continue it analytically}. With this notation the residue term is $e^{-ip_{1}L}N_{\mathrm{sum}}(\theta_{2},\theta_{1}
+3i\pi)$ and as we explained 
$N_{\mathrm{sum}}(\theta_{2},\theta_{1}+3i\pi)$ now denotes the sum
\begin{eqnarray}
 &  & O^c(\theta_{2},\theta_{1}+3i\pi)+\int_{-\infty}^{\infty}\frac{du}{2\pi}\mu_{1}(u)O(\theta_{2},\theta_{1}+3i\pi,u^+,u^-)e^{-LE(u)}+\\
 &  &\frac{1}{2}\int_{-\infty}^{\infty}\frac{du_{1}}{2\pi}\int_{-\infty}^{\infty}\frac{du_{2}}{2\pi}\,\mu_{2}(u_{1},u_{2})O^c(\theta_{2},,\theta_{1}+3i\pi,u_{1}^+,u_{2}^+,u_{2}^-,u_{1}^-)e^{-L\left(E(u_{1})+E(u_{2})\right)}
 +\dots \nonumber 
\end{eqnarray}
For the DSFT axioms to be fulfilled in the general case it is crucial that the factor $e^{-ip_{1}L}$ can be factored out from 
each term. This can be guaranteed by demanding
\begin{equation}
-i\mbox{res}_{\theta_{1}+i\pi =u_{1}}\mu_{2}(u_{1},u_2)O^c(\theta_{1}+i\pi ,\theta_{2},u_{1}^+,u_{1}^-,u_{2}^+,u_{2}^-)=  
\mu_{1}(u_{2})O^c(\theta_{2},\theta_{1}+3i\pi ,u_{2}^+,u_{2}^-)
\end{equation}
and assuming that higher order poles do not contribute. The mechanism producing the extra residue term was called teleportation in
\cite{HEXAGON}. This also indicates, how we should split the particles between
regions \#3 and \#3': we should sum for all possible distributions
with an additional $e^{-ipL}$ factor, whenever a particle is moved
to region \#3'. Observe that it is crucial that we do not have any
contributions from double or higher order integrations as they would
spoil the above structure. Actually in the free boson case we know that there is a double pole contribution, which can be 
compensated by an appropriately chosen measure factor. Thus the existence of higher order poles leads to non-trivial measure 
factors to guarantee
\begin{equation}
N_L(\theta_1 +i\pi ,\theta _2)=N_{\mathrm{sum}}(\theta_{1}+i\pi ,\theta_{2})+
e^{-ip_{1}L}N_{\mathrm{sum}}(\theta_{2},\theta_{1}+3i\pi)
\end{equation}
This the equation we should satisfy in the general interacting case. Demanding it for the continued rapidities will give restrictions on the definition 
of the connected octagon form factors and the measure. 

Now, assuming that $O^c(\theta_{1},\theta_{2},u_{1}^+,\dots,u_{1}^-)$
is non-singular at $\theta_{1}+i\pi =\theta_{2}$ (actually it is
true in the free boson theory as $k(\theta+i\pi)=-k(\theta)$ and
follows from our normalization $N_{L}(\emptyset)=1$ in general) we
obtain the kinematical singularity equation 
\begin{equation}
-i\mbox{res}_{\theta'=\theta}N_{L}(\theta'+i\pi,\theta)=(1-e^{-ipL})
\end{equation}
\begin{figure}[H]
\centering{}\includegraphics[height=3.5cm]{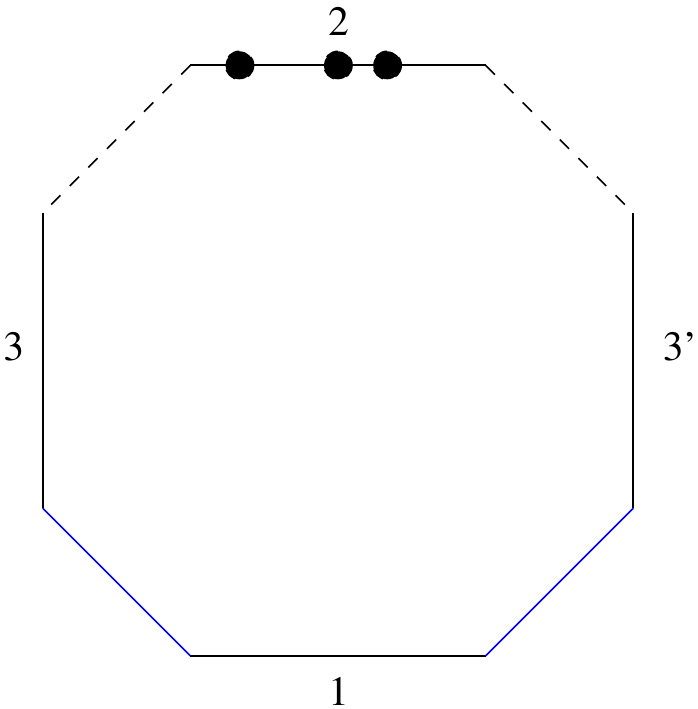} $\to$\includegraphics[height=3.5cm]{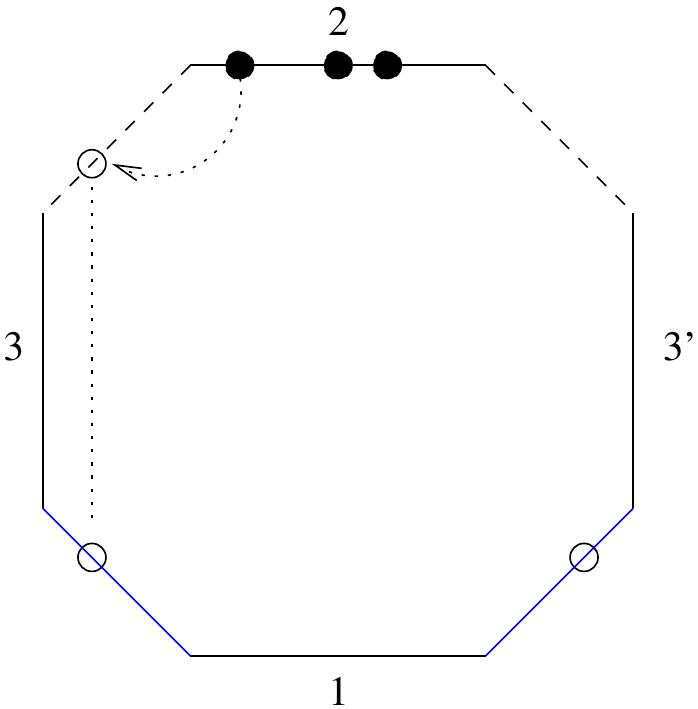}$\to$
\includegraphics[height=3.5cm]{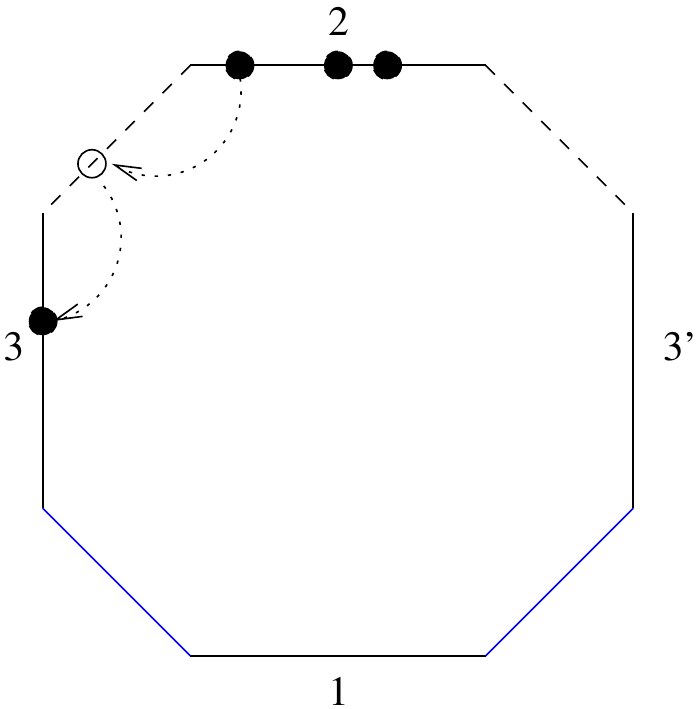}$+$ $e^{-ipL}$\includegraphics[height=3.5cm]{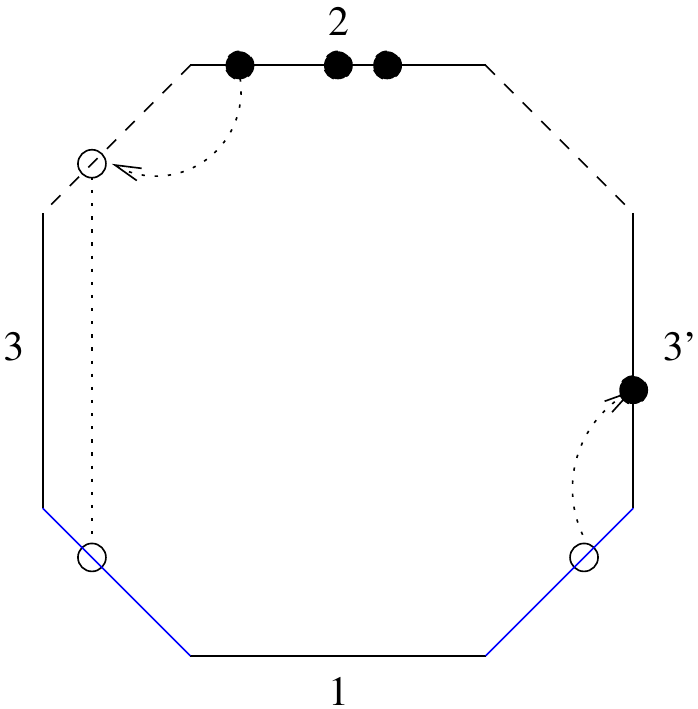}\caption{Analytical continuation from domain \#2 to domain \#3. }
\label{cont1}
\end{figure}

\item In order to show the periodicity axiom we need to continue further
$\theta_{1}+i\pi \to\theta_{1}+2i\pi $. In doing the continuation 
in each term of the sum $N_{\mathrm{sum}}(\theta_{1}+i\pi ,\theta_{2})$ we do not
expect any teleportation as the $\theta_{1}^++i\pi =u_{1}^-$
singularity is regularized in the connected part. In continuing in terms of the sum $N_{\mathrm{sum}}(\theta_{2},\theta_{1}+3i\pi )$
we expect both the direct and the teleported terms, such that the teleported residue term is proportional to the direct term,
see Figure \ref{cont2}:
\begin{eqnarray}
N_L(\theta_1 +2i\pi ,\theta _2)=&&N_{\mathrm{sum}}(\theta_{1}+2i\pi ,\theta_{2})+ 
e^{ip_{1}L}N_{\mathrm{sum}}(\theta_{2},\theta_{1}+4i\pi) \nonumber \\ && + e^{ip_{1}L}(-e^{-ip_1L})
N_{\mathrm{sum}}(\theta_{1}+2i\pi ,\theta_{2})
\end{eqnarray}

\begin{figure}[H]
\begin{centering}
~\includegraphics[height=3.5cm]{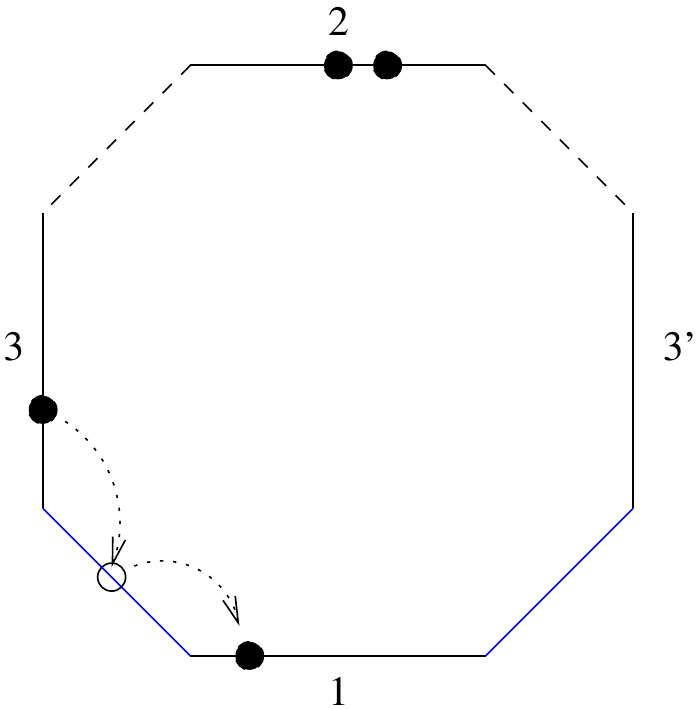}$+e^{ipL}(-e^{-ipL})$
\includegraphics[height=3.5cm]{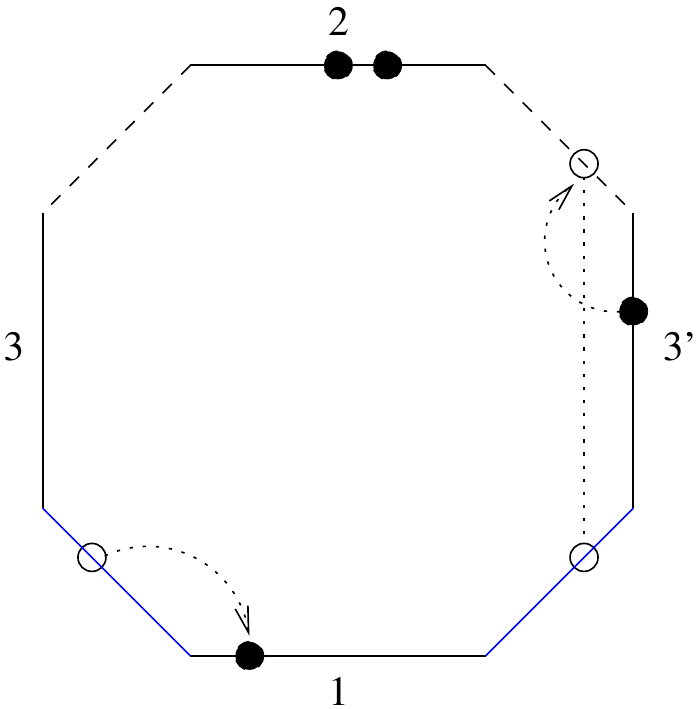}$+$$e^{ipL}$\includegraphics[height=3.5cm]{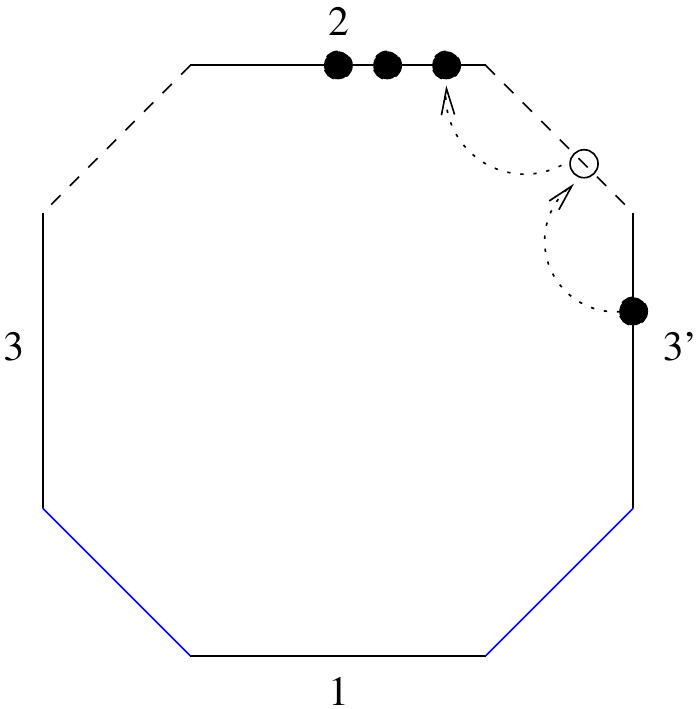}
\par\end{centering}
\caption{Analytical continuation from domain \#3 and \#3' back to domain \#2. }
\label{cont2}
\end{figure}

The direct continuation from $N_{\mathrm{sum}}(\theta_{1}+i\pi ,\theta_{2})$
and the teleported continuation from $N_{\mathrm{sum}}(\theta_{2},\theta_{1}+3i\pi )$
cancel each other and only the direct continuation from $N_{\mathrm{sum}}(\theta_{2},\theta_{1}+3i\pi )$
remains. This result is precisely the expected relation 
\begin{equation}
N_{L}(\theta_{1}+2i\pi,\theta_{2})=e^{ip_{1}L}N_{L}(\theta_{2},\theta_{1})
\end{equation}
\end{itemize}


\begin{thebibliography}{[99]}


\bibitem{adscft1}
 J.~M.~Maldacena,
  ``The Large N limit of superconformal field theories and supergravity,''
  Int.\ J.\ Theor.\ Phys.\  {\bf 38} (1999) 1113
   [Adv.\ Theor.\ Math.\ Phys.\  {\bf 2} (1998) 231]
  [hep-th/9711200].

\bibitem{Bena:2003wd}
  I.~Bena, J.~Polchinski and R.~Roiban,
  ``Hidden symmetries of the AdS(5) x S**5 superstring,''
  Phys.\ Rev.\ D {\bf 69} (2004) 046002
  [hep-th/0305116].

\bibitem{Arutyunov:2008if}
  G.~Arutyunov and S.~Frolov,
  ``Superstrings on AdS(4) x CP**3 as a Coset Sigma-model,''
  JHEP {\bf 0809} (2008) 129
  [arXiv:0806.4940 [hep-th]].

\bibitem{Stefanski:2008ik}
  B.~Stefanski, jr,
  ``Green-Schwarz action for Type IIA strings on AdS(4) x CP**3,''
  Nucl.\ Phys.\ B {\bf 808} (2009) 80
  [arXiv:0806.4948 [hep-th]].

\bibitem{Babichenko:2009dk}
  A.~Babichenko, B.~Stefanski, Jr. and K.~Zarembo,
  ``Integrability and the AdS(3)/CFT(2) correspondence,''
  JHEP {\bf 1003} (2010) 058
  [arXiv:0912.1723 [hep-th]].


  \bibitem{intreview}
For a review see the collection
  N.~Beisert, C.~Ahn, L.~F.~Alday, Z.~Bajnok, J.~M.~Drummond, L.~Freyhult, N.~Gromov and R.~A.~Janik {\it et al.},
  ``Review of AdS/CFT Integrability: An Overview,''
  Lett.\ Math.\ Phys.\  {\bf 99} (2012) 3
  [arXiv:1012.3982 [hep-th]].

  
\bibitem{TBA1}
  G.~Arutyunov and S.~Frolov,
  ``String hypothesis for the AdS(5) x S**5 mirror,''
  JHEP {\bf 0903} (2009) 152
  [arXiv:0901.1417 [hep-th]].

\bibitem{TBA2}
  N.~Gromov, V.~Kazakov and P.~Vieira,
  ``Exact Spectrum of Anomalous Dimensions of Planar N=4 Supersymmetric Yang-Mills Theory,''
  Phys.\ Rev.\ Lett.\  {\bf 103} (2009) 131601
  [arXiv:0901.3753 [hep-th]].

\bibitem{TBA3}
  D.~Bombardelli, D.~Fioravanti and R.~Tateo,
  ``Thermodynamic Bethe Ansatz for planar AdS/CFT: A Proposal,''
  J.\ Phys.\ A {\bf 42} (2009) 375401
  [arXiv:0902.3930 [hep-th]].

\bibitem{BH}
  J.~Balog and A.~Hegedus,
  ``Hybrid-NLIE for the AdS/CFT spectral problem,''
  JHEP {\bf 1208} (2012) 022
  [arXiv:1202.3244 [hep-th]].


\bibitem{QSC1}
 N.~Gromov, V.~Kazakov, S.~Leurent and D.~Volin,
  ``Quantum Spectral Curve for Planar $\mathcal{N} =$ Super-Yang-Mills Theory,''
  Phys.\ Rev.\ Lett.\  {\bf 112} (2014) 1,  011602
  [arXiv:1305.1939 [hep-th]].

\bibitem{QSC2}
  N.~Gromov, V.~Kazakov, S.~Leurent and D.~Volin,
  ``Quantum spectral curve for arbitrary state/operator in AdS$_{5}$/CFT$_{4}$,''
  JHEP {\bf 1509} (2015) 187
  [arXiv:1405.4857 [hep-th]].

\bibitem{SFT}
  Z.~Bajnok and R.~A.~Janik,
  ``String field theory vertex from integrability,''
  JHEP {\bf 1504} (2015) 042
  [arXiv:1501.04533 [hep-th]].

\bibitem{Bajnok:2014sza}
  Z.~Bajnok, R.~A.~Janik and A.~Wereszczyński,
  ``HHL correlators, orbit averaging and form factors,''
  JHEP {\bf 1409} (2014) 050
  [arXiv:1404.4556 [hep-th]].

\bibitem{Hollo:2015cda}
  L.~Hollo, Y.~Jiang and A.~Petrovskii,
  ``Diagonal Form Factors and Heavy-Heavy-Light Three-Point Functions at Weak Coupling,''
  JHEP {\bf 1509} (2015) 125
  [arXiv:1504.07133 [hep-th]].
 
\bibitem{Bajnok:2016xxu}
  Z.~Bajnok and R.~A.~Janik,
  ``Classical limit of diagonal form factors and HHL correlators,''
  JHEP {\bf 1701} (2017) 063
  [arXiv:1607.02830 [hep-th]].

\bibitem{HEXAGON} 
  B.~Basso, S.~Komatsu and P.~Vieira,
  ``Structure Constants and Integrable Bootstrap in Planar N=4 SYM Theory,''
  arXiv:1505.06745 [hep-th].


\bibitem{Eden:2015ija}
  B.~Eden and A.~Sfondrini,
  ``Three-point functions in ${\cal N}=4$ SYM: the hexagon proposal at three loops,''
  JHEP {\bf 1602} (2016) 165
  [arXiv:1510.01242 [hep-th]].

\bibitem{Basso:2015eqa}
  B.~Basso, V.~Goncalves, S.~Komatsu and P.~Vieira,
 ``Gluing Hexagons at Three Loops,''
  Nucl.\ Phys.\ B {\bf 907} (2016) 695
  [arXiv:1510.01683 [hep-th]].

\bibitem{Jiang:2016ulr}
  Y.~Jiang, S.~Komatsu, I.~Kostov and D.~Serban,
  ``Clustering and the Three-Point Function,''
  J.\ Phys.\ A {\bf 49} (2016) no.45,  454003
  [arXiv:1604.03575 [hep-th]].

\bibitem{Jiang:2015bvm}
  Y.~Jiang and A.~Petrovskii,
  ``Diagonal form factors and hexagon form factors,''
  JHEP {\bf 1607} (2016) 120
  [arXiv:1511.06199 [hep-th]].

\bibitem{Jiang:2016dsr}
  Y.~Jiang,
  ``Diagonal Form Factors and Hexagon Form Factors II. Non-BPS Light Operator,''
  JHEP {\bf 1701} (2017) 021
  [arXiv:1601.06926 [hep-th]].

  
\bibitem{Basso:2017muf}
  B.~Basso, V.~Goncalves and S.~Komatsu,
  ``Structure constants at wrapping order,''
  arXiv:1702.02154 [hep-th].
  


  


\bibitem{KINEMATICAL}
  Z.~Bajnok and R.~A.~Janik,
  ``The kinematical AdS$_{5} \times$ S$^{5}$ Neumann coefficient,''
  JHEP {\bf 1602} (2016) 138
  [arXiv:1512.01471 [hep-th]].  

\bibitem{Zamolodchikov:1978xm}
  A.~B.~Zamolodchikov and A.~B.~Zamolodchikov,
  ``Factorized s Matrices in Two-Dimensions as the Exact Solutions of Certain Relativistic Quantum Field Models,''
  Annals Phys.\  {\bf 120} (1979) 253.

\bibitem{Mussardo:1992uc}
  G.~Mussardo,
  ``Off critical statistical models: Factorized scattering theories and bootstrap program,''
  Phys.\ Rept.\  {\bf 218} (1992) 215.

\bibitem{KW}
  M.~Karowski and P.~Weisz,
  ``Exact Form-Factors in (1+1)-Dimensional Field Theoretic Models with Soliton Behavior,''
  Nucl.\ Phys.\ B {\bf 139}, 455 (1978).

\bibitem{Smirnov} 
  F.~A.~Smirnov,
  ``Form-factors in completely integrable models of quantum field theory,''
  Adv.\ Ser.\ Math.\ Phys.\  {\bf 14}, 1 (1992).

\bibitem{Pozsgay:2007kn}
  B.~Pozsgay and G.~Takacs,
  ``Form-factors in finite volume I: Form-factor bootstrap and truncated conformal space,''
  Nucl.\ Phys.\ B {\bf 788} (2008) 167
  [arXiv:0706.1445 [hep-th]].

\bibitem{Pozsgay:2007gx}
  B.~Pozsgay and G.~Takacs,
  ``Form factors in finite volume. II. Disconnected terms and finite temperature correlators,''
  Nucl.\ Phys.\ B {\bf 788} (2008) 209
  [arXiv:0706.3605 [hep-th]].


\bibitem{LUSCHER}
  M.~Luscher,
  ``Volume Dependence of the Energy Spectrum in Massive Quantum Field Theories. 1. Stable Particle States,''
  Commun.\ Math.\ Phys.\  {\bf 104} (1986) 177.

\bibitem{KONISHI}
  Z.~Bajnok and R.~A.~Janik,
  ``Four-loop perturbative Konishi from strings and finite size effects for multiparticle states,''
  Nucl.\ Phys.\ B {\bf 807} (2009) 625
  [arXiv:0807.0399 [hep-th]].

\bibitem{BOMBARDELLI}
  D.~Bombardelli,
  ``A next-to-leading L\"uscher formula,''
  JHEP {\bf 1401} (2014) 037
  [arXiv:1309.4083 [hep-th]].

  \bibitem{Zamolodchikov} 
  A.~B.~Zamolodchikov,
  ``Thermodynamic Bethe Ansatz in Relativistic Models. Scaling Three State Potts and Lee-yang Models,''
  Nucl.\ Phys.\ B {\bf 342}, 695 (1990).
  


\bibitem{LSNS}
  J.~Lucietti, S.~Schafer-Nameki and A.~Sinha,
  ``On the plane wave cubic vertex,''
  Phys.\ Rev.\ D {\bf 70} (2004) 026005
  [hep-th/0402185].

 

\bibitem{Leclair:1999ys}
  A.~Leclair and G.~Mussardo,
  ``Finite temperature correlation functions in integrable QFT,''
  Nucl.\ Phys.\ B {\bf 552} (1999) 624
  [hep-th/9902075].


\bibitem{Pozsgay:2010xd}
  B.~Pozsgay,
  ``Mean values of local operators in highly excited Bethe states,''
  J.\ Stat.\ Mech.\  {\bf 1101} (2011) P01011
  [arXiv:1009.4662 [hep-th]].
  

\bibitem{Basso:2017khq}
  B.~Basso, F.~Coronado, S.~Komatsu, H.~T.~Lam, P.~Vieira and D.~l.~Zhong,
  ``Asymptotic Four Point Functions,''
  arXiv:1701.04462 [hep-th].

\bibitem{Bargheer:2017eoz}
  T.~Bargheer,
  ``Four-Point Functions with a Twist,''
  arXiv:1701.04424 [hep-th].

\bibitem{Fleury:2016ykk}
  T.~Fleury and S.~Komatsu,
  ``Hexagonalization of Correlation Functions,''
  JHEP {\bf 1701} (2017) 130
  [arXiv:1611.05577 [hep-th]].
 
\bibitem{Eden:2016xvg}
  B.~Eden and A.~Sfondrini,
  ``Tessellating cushions: four-point functions in N=4 SYM,''
  arXiv:1611.05436 [hep-th].


  
  
\end{thebibliography}
\end{document}